\documentclass[longbibliography,
twocolumn,
showpacs,
floatfix,
aps,
prb,
amsmath,
amssymb,
superscriptaddress,
10pt]{revtex4-2}

\usepackage[utf8]{inputenc} % UTF-8 encoding
\usepackage[T1]{fontenc}    % Font encoding
\usepackage{amsmath}        % Advanced math
\usepackage{amssymb}        % Math symbols
\usepackage{graphicx}      % per le immagini
\usepackage{hyperref}       % Clickable links
\usepackage{physics}        % Physics symbols
\usepackage{xcolor}         % Custom colors
\usepackage{enumitem}       % Custom lists
\usepackage{bm}             % Bold math symbols
\usepackage{float}          % Advanced float placement
\usepackage{makecell}       % make better cells for tables
\usepackage[english]{babel} %Language package

\usepackage{verbatim}
\usepackage{tikz-feynman}
\usepackage{amsmath}
\usepackage[capitalize]{cleveref}
\usepackage[normalem]{ulem} %for striking out text with \sout{} (useful for corrections)
\usepackage{cancel} %for striking out text with \cancel{} (useful for corrections)
\usepackage{orcidlink} %for adding orcid links

\tikzfeynmanset{warn luatex=false, compat=1.1.0}

\usetikzlibrary{decorations.pathmorphing, arrows.meta, positioning}
\tikzset{
    photon/.style={decorate, decoration={snake, amplitude=2pt, segment length=4pt}, draw},
    fermion/.style={draw=black, postaction={decorate}, decoration={markings, mark=at position 0.5 with {\arrow{>}}}}
}

\hypersetup{
    colorlinks=true,
    linkcolor=blue,
    citecolor=blue,
    filecolor=magenta,
    urlcolor=cyan,
}

\definecolor{igcolor}{RGB}{0,120,100}

\begin{document}

\title{Formation of Cavity-Polaritons via High-Order Van Hove Singularities}
\author{Igor Gianardi~\orcidlink{0009-0003-8788-1823}}
\email{gianardi@pks.mpg.de}
\affiliation{Max Planck Institute for the Physics of Complex Systems, 01187 Dresden, Germany}

\author{Michele Pini~\orcidlink{0000-0001-5522-5109}}
\email{michele.pini@uni-a.de}
\affiliation{Theoretical Physics III, Center for Electronic Correlations and Magnetism,
Institute of Physics, University of Augsburg, 86135 Augsburg, Germany}
\affiliation{Max Planck Institute for the Physics of Complex Systems, 01187 Dresden, Germany}

\author{Francesco Piazza~\orcidlink{0000-0003-1332-6627}}
\affiliation{Theoretical Physics III, Center for Electronic Correlations and Magnetism,
Institute of Physics, University of Augsburg, 86135 Augsburg, Germany}
\affiliation{Max Planck Institute for the Physics of Complex Systems, 01187 Dresden, Germany}

\date{\today}

\begin{abstract} 
We consider polaritons formed by hybridizing a continuum of interband  particle-hole excitations of an insulating phase with a cavity photon at subgap frequencies, where absorption is suppressed. 
The strength of the hybridization is driven by the Van Hove singularity in the joint density of states (JDOS) at the band gap: the stronger the singularity, the more a photon is hybridized with the interband transitions. 
In order to increase the singularity and thus the polariton hybridization without absorption, we propose to engineer a nonparabolic momentum dispersion of the bands around the gap in order to implement a high-order Van Hove singularity (HOVHS) in the JDOS. Ultracold atoms in tunable optical lattices are an ideal platform to engineer two-dimensional gapped phases with nontrivial band dispersions at the gap. Moreover, the intrinsic noninteracting nature of polarized fermionic atoms prevents the emergence of  subgap excitations, which are common in solid-state systems and could otherwise spoil the absence of absorption below the gap. 
Our findings identify band-engineering at the gap edge as a promising route for polariton control with applications in quantum nonlinear optics. 
\end{abstract}

\maketitle

\section{Introduction}\label{sec:Intro}
Polaritons are hybrid quasi-particles that blend matter and light properties \cite{basov2020polariton}: an excitation of the electromagnetic field (i.e.~a photon) can be converted into an excitation of the matter, and vice versa. For applications of polaritonics in quantum (nonlinear) optics \cite{chang2014quantum} it is crucial that this back-and-forth exchange of excitations between the electromagnetic field and the matter takes place efficiently. This requires two main features: i) the coupling between light and matter to be sufficiently strong, and ii) the absorption of photons from the matter to be sufficiently small compared to the energy shift caused by the matter. The use of cavities for light confinement is useful for both i) and ii), as it increases the light-matter coupling as well as allows for tuning the photon spectrum with respect to the matter excitations.

The matter excitations most commonly considered for cavity-polariton formation appear as discrete resonances in the optical response of the uncoupled material. A preexisting discrete material resonance is not, however, essential. Indeed, coupling the cavity mode to a continuum of matter excitations can generate a discrete polariton state below the continuum’s threshold.
 In this spectral region, a photon can experience a large energy shift through hybridization with virtual continuum excitations without being absorbed by the continuum.

An example is provided by Fermi-edge polaritons, observed in doped quantum wells~\cite{Gabbay2007}, where the cavity couples to an interband particle-hole continuum whose threshold marks the onset of transitions from the valence band to unoccupied conduction-band states above the Fermi level~\cite{Pimenov2017}.
Polaritons have also been predicted~\cite{Cortese2019} and observed~\cite{Cortese2021} in narrow doped quantum wells via transitions from the single bound conduction subband to unbound states, with the continuum threshold set by the ionization energy~\cite{Kumar2022}. Their recently predicted strong interactions make them promising for (quantum) nonlinear optics~\cite{OliveiraCiuti2025}.
The same mechanism of polariton formation has been observed in insulating bilayer graphene coupled to a subwavelength Terahertz resonator, where the particle-hole continuum instead arises from transitions between the valence and conduction bands, with its lower edge set by the band-gap energy~\cite{Helmrich2026}.

In this work, we focus on the interband particle-hole continuum of an insulating material. We explore the possibility to enhance light–matter hybridization by shaping the structure of the continuum to
increase the number of possible states between which the particle-hole can be excited, which is quantified by the joint density of states (JDOS). In particular, the JDOS can even diverge at a Van Hove singularity (VHS). 
Since the divergence strength of VHSs increases as the system's dimensionality is reduced, one-dimensional systems are the most obvious candidates for the implementation of this mechanism. For such systems, the singularity consists of an inverse square-root divergence~\cite{Razeghi2018,grosso2013solid}. Indeed, the emergence of a cavity-polariton in presence of a VHS of an insulating phase was considered in the context of 1D nanowires~\cite{Arnardottir2013}.
Here, we explore a promising alternative route to lowering the dimensionality  to enhance the hybridization of Van Hove polaritons, namely the engineering of a nonparabolic momentum dispersion of the bands around the gap in order to implement a high-order Van Hove singularity (HOVHS) in the JDOS. 

We recall that a VHS is a nonanalytic feature of the density of states associated with a critical point of the energy dispersion in the band structure. 
Considering for simplicity a single-band dispersion \(E(\bm{k})\), a quasi-momentum \(\bm{k}^*\) is a critical point if the group velocity vanishes there, i.~e.~\(
\nabla_{\bm{k}}E(\bm{k})|_{\bm{k}=\bm{k}^*}=\bm{0}.
\)
The local structure of the dispersion near \(\bm{k}^*\) is obtained from its Taylor expansion in the displacement \(\bm{k}-\bm{k}^*\). Since the linear term vanishes at a critical point, the first possible non-constant contribution is governed by the Hessian matrix \(\mathcal{H}_{\mu\nu}(\bm{k}^*)=\partial_{k_\mu}\partial_{k_\nu}E(\bm{k})|_{\bm{k}=\bm{k}^*}\), namely the matrix of second derivatives with respect to quasi-momentum. The VHS is of ordinary type when \(\det\mathcal{H}(\bm{k}^*)\neq0\), i.e.~when the leading local variation of the band energy is quadratic in all directions. A HOVHS is instead associated with a degenerate critical point, \(\det\mathcal{H}(\bm{k}^*)=0\), meaning that at least one eigenvalue of the Hessian vanishes. The leading dispersion along the corresponding direction is then controlled by higher-order terms in the Taylor expansion, resulting in a stronger power-law divergence of the density of states (DOS). If one considers a gapped insulator, the same reasoning applies, but with the single-band dispersion \(E(\bm{k})\) replaced by the interband transition energy
\(
\delta E_{\bm{k}}=E_{+}(\bm{k})-E_{-}(\bm{k}),
\)
where \(E_{-}(\bm{k})\) and \(E_{+}(\bm{k})\) denote the valence and conduction band dispersions, respectively. In this case, one speaks of a VHS or HOVHS in the JDOS of interband excitations, rather than in the DOS of a single band.

HOVHSs have recently attracted interest in the solid-state community as a mean to modify transport and thermodynamical quantities, as well as enhance electron correlations~\cite{efremov_multicritical_2019, Yuan2019HighOrderVHS,Class_HOVHS, classen_competing_2020,Graphene_VHS_overdoping_2020, Kagome_2022, seiler_quantum_2022, chandrasekaran_engineering_2024,classen_high-order_2025,Burke2025}. In our case, we suggest instead to employ a HOVHS to increase the light-matter hybridization.
In particular, we identify a specific band shape of a checkerboard lattice that gives rise to a new type of HOVHS, characterized by an inverse-square root divergence of the JDOS, further enhanced by a logarithm. To the best of our knowledge, this kind of HOVHS has not yet been reported in the existing solid-state-physics literature~\cite{Class_HOVHS} and turns out to be useful for polaritonics.

While solid-state materials with 2D checkerboard lattices can be in principle realized \cite{Hu2023}, we envision the cleanest implementation of high-order van Hove polaritons to be achievable with an ultracold gas of polarized fermionic atoms trapped in a checkerboard optical lattice and dispersively coupled to a single-mode optical cavity—a state-of-the-art setup \cite{Mivehvar2021}. The advantage of ultracold atoms over solid-state materials is the enhanced possibilities of band engineering, enabled by the tunability of the laser interference pattern \cite{InguscioKetterleSalomon2007,Giorgini2008,BlochDalibardZwerger2008,Li2016}. Moreover, the form and strength of the light-matter coupling has been shown to be widely controllable \cite{Mivehvar2021}, such that a nonzero coupling can be engineered for the relevant energies at which the HOVHS is present. Finally and probably most crucially, in ultracold atomic platforms subgap excitations that would introduce absorption and spoil the VHS are typically absent. Optical lattices are rigid and impurity-free, and the interatomic interactions can be tuned to zero e.g.~using polarized fermionic atoms. 
Indeed, while our description applies in principle also to solid-state platforms, achieving negligible absorption below the band gap is generally more challenging in these systems than in ultracold atom platforms. This is due to the presence of electron-electron interactions or additional degrees of freedom, such as phonons.
In experiments, the difficulty to observe signatures of the VHS has been for instance attributed to excitons -- observed in nanowires~\cite{GaAs_Room_2005,Zno_Pol_Las_2002}, carbon nanotubes~\cite{Carbon_nano_2007}, conjugated polymers~\cite{Rohlfing1999, Rohlfing2000}, and bilayer graphene~\cite{TunableExcitons} -- as well as more exotic types of excitations, as found in Blue-Bronzes~\cite{Degiorgi1991}.
Nevertheless, signatures of a VHS have been reported in a two-dimensional topological insulator~\cite{Jiang2020VHS}, and recent experiments on insulating bilayer graphene coupled to a subwavelength Terahertz resonator have even revealed subgap polaritons, with no discrete excitation detected at these energies in the uncoupled material~\cite{Helmrich2026}, opening up the possibility of testing our predictions in solid-state systems.

The paper is structured as follows.
Section~\ref{sec:Key_Results} presents our main results, showing how a novel 2D HOVHS can strongly enhance the hybridization of Van Hove polaritons. In Section~\ref{sec:Van_Hove_Pol}, we show the general mechanism behind the emergence of this class of polaritons. The appearance of the HOVHS in a 2D checkerboard lattice model is presented in Section~\ref{sec:cVHS}, and its implementation with ultracold atoms is discussed in Section~\ref{sec:Ultra-imp}.
Throughout the paper, we set the reduced Planck constant $\hbar$ to unity with the exception of \cref{ssec:Feasibility}, where the value of $\hbar$ is restored to provide an estimate of experimental parameters.
\section{Key Results}\label{sec:Key_Results}
\begin{figure}[tbp]
    \centering   \includegraphics[width=0.5\textwidth]{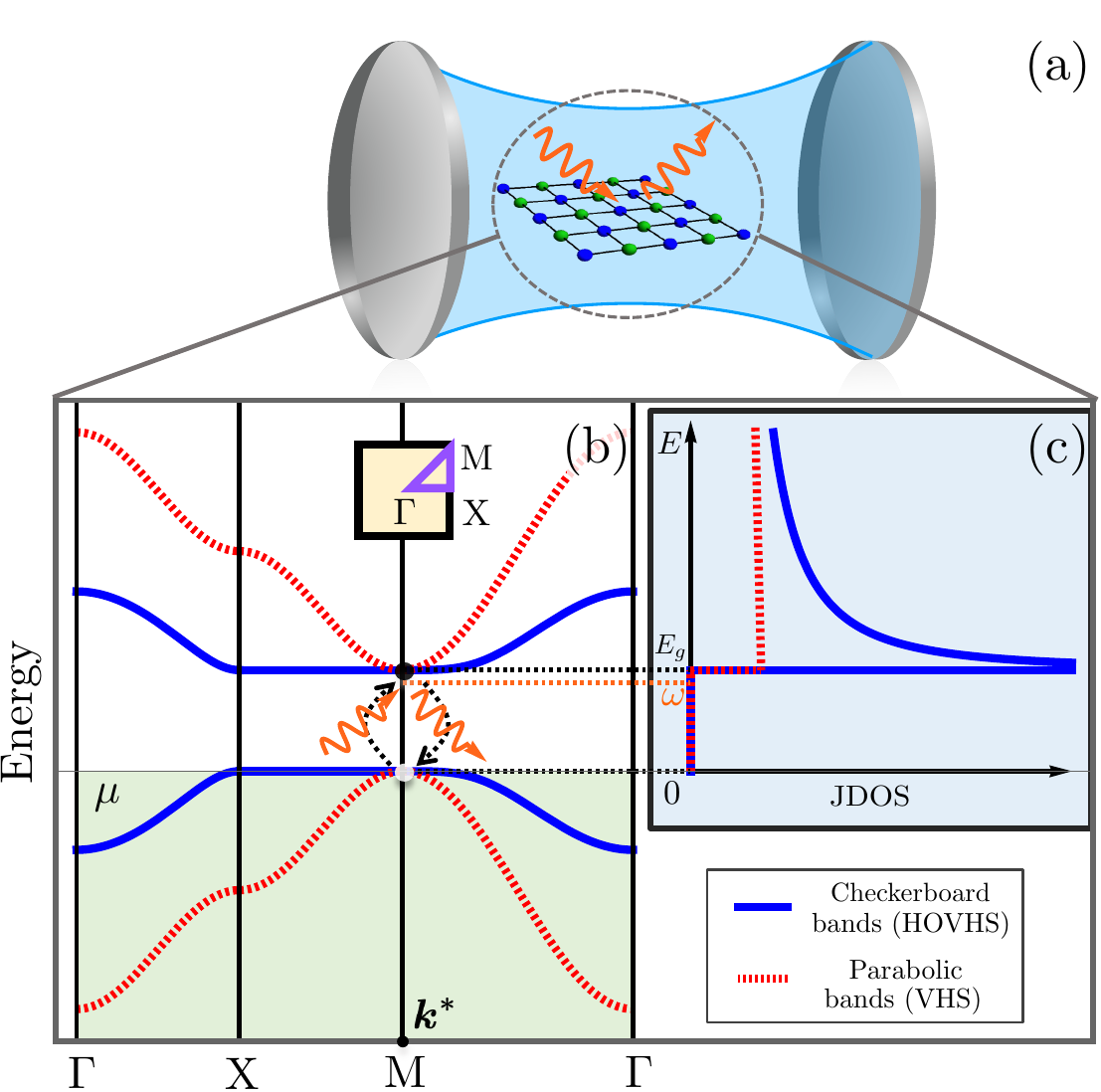}
  \caption{
Schematic illustration of the polariton formation mechanism at the gap edge of an insulator with engineered band dispersion.
(a) Sketch of the system in real space: a material with a checkerboard lattice embedded in an optical cavity. (b) Energy bands of the checkerboard material (blue continuous line) compared to those of a standard parabolic insulator (red dotted line).
The $\text{M}$ point coincides in both cases with the critical point $\bm{k}^*$ of the Van Hove singularity (VHS). The chemical potential is chosen so that the valence band is completely filled (green shaded region) and the conduction band is empty (white region), in order to have an insulating phase. Cavity photons (orange wavy arrows) with energy slightly below the band gap ($ \omega \lesssim E_g$) hybridize with interband particle-hole excitations (black dashed arrows and circles). (c) Joint density of states (JDOS) for both insulators. The parabolic bands show an ordinary step-like VHS at the gap, while the engineered checkerboard bands feature a high-order VHS (HOVHS).
}
\label{fig:sketch}
\end{figure}
The scenario considered here is illustrated in  Fig.~\ref{fig:sketch}. Photons confined within a cavity are coupled to particle-hole excitations of a crystalline material [see panel (a)]. These excitations correspond to interband transitions in the band structure depicted in panel (b).
As explained in the introduction, in order to maximize energy shifts by keeping the absorption negligible, we consider the case where the density of states associated with the interband excitations, namely the JDOS, is large above a gap of size \(E_g\) and vanishes immediately below it, as shown for example by the blue solid curve in panel (c) of Fig.~\ref{fig:sketch}. 
Therefore, a photon with frequency \(\omega \lesssim E_g\) interacting with vertical interband particle-hole excitations at quasi-momentum $\bm{k}$ 
with a light-matter coupling strength \(g(\bm{k})\) simultaneously experiences no absorption (as it cannot excite a particle into the conduction band) and a finite energy shift due to hybridization with the virtual  excitations.
As anticipated in the introduction, in order to increase the energy shift experienced by the photon without increasing the absorption, we consider to increase the JDOS of the excitations in the material. Instead of reducing the dimensionality of the latter, we propose to tailor the shape of the bands around the gap to increase the order of the VHS in the JDOS,  implementing a HOVHS.
With this goal in mind, we identify a band-insulating phase on a checkerboard lattice, which gives rise to a new type of HOVHS, corresponding to a $\log(\omega-E_g)/\sqrt{\omega-E_g}$ divergence in the JDOS at the band gap [see \cref{fig:sketch}(c)]. 
 In this case, the critical quasi-momentum \(\bm{k}^*\) coincides with the \(\mathrm{M}\) point of the Brillouin zone (BZ). The HOVHS is generated by the dispersion of the bands near this point, which is quartic along the $\Gamma\text{M}$ direction within the BZ, and remains flat along its edges, as illustrated in  panel (b) of \cref{fig:sketch} (blue solid line). This deviates from the standard 2D parabolic case (red dotted line), which instead presents only an ordinary step-like VHS in the JDOS at the gap.
\begin{figure}[tbp]
\centering
\includegraphics[width=\linewidth]{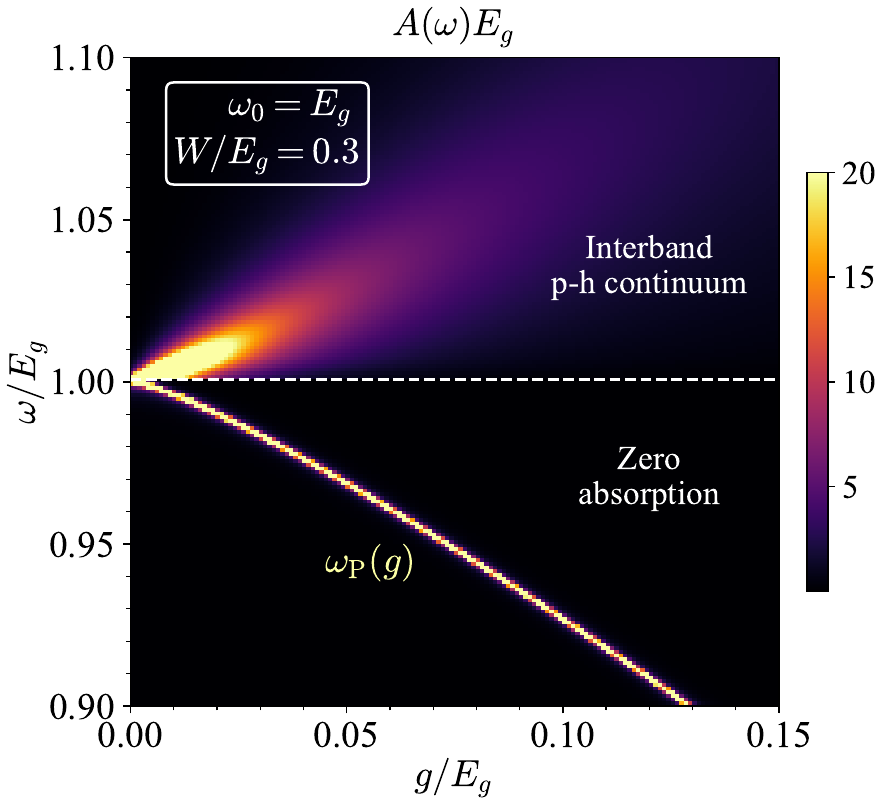}
\caption{Photon spectral function \(A(\omega)\) as a function of photon energy \(\omega\) and coupling strength \(g\), with the bare cavity frequency \(\omega_0\) set equal to the band gap \(E_g\) of a checkerboard 2D lattice. The spectrum reveals two distinct regions. For \(\omega > E_g\), the spectral function exhibits broadened features dominated by absorption processes. For \(\omega < E_g\), a lower polaritonic branch emerges, characterized by zero absorption and a large energy shift. 
The figure is obtained with a representative rate of cavity losses $\kappa/E_g=10^{-4}.$} \label{fig:densityplot}
\end{figure}

The formation of polaritons in this setting is shown in Fig.~\ref{fig:densityplot}, where the photon spectral function \(A(\omega)\) (defined in Section \ref{sec:Van_Hove_Pol}) is plotted as a function of the photon frequency \(\omega\) and the coupling strength \( g \equiv g(\bm{k}^{\ast})
\), where we assumed the $\bm{k}$-dependence of $g(\bm{k})$ to be negligible near the critical point $\bm{k}^*$. Here, the bare cavity frequency \(\omega_0\) is set to be resonant with interband transitions at the band gap \(E_g\) in order to maximize the hybridization.
In evaluating the spectral function, we included a finite cavity linewidth \(\kappa\), which accounts for photon losses and sets the intrinsic spectral width of the photon-like component of the polariton.
Moreover, we denote by \(W\) the single-particle bandwidth of the relevant insulating bands, namely the energy range over which either the valence or conduction band disperses away from the gap edge. Together with the gap \(E_g\), \(W\) fixes the energy window of vertical interband transitions: the particle-hole continuum extends from \(E_g\) to \(E_g+2W\). The dimensionless ratio \(W/E_g\) therefore measures how broad the bands are compared with the gap. In a tight-binding realization, \(W\) is typically controlled by the hopping amplitude \(t\) at fixed \(E_g\). For the checkerboard model introduced below, its explicit dependence on \(t\) and \(E_g\) is given in \cref{eqn:checkerboard_bandwidth}.
When the photon has enough energy to excite the particle-hole continuum above the gap (\(\omega > E_g\)), the spectral function shows a broadened polariton branch dominated by absorption processes. In this region, the spectral width can be understood
as the sum of the cavity-loss contribution, set by the finite linewidth \(\kappa\), and
the absorption rate of the matter excitations in the particle-hole continuum. We do not
investigate this branch further, since its absorption-dominated character makes it less
relevant for the purposes of quantum optics. For photon frequencies below the gap (\(\omega < E_g\)), the spectral function exhibits instead a very narrow, i.e.~long-lived polaritonic branch $\omega_{\mathrm{P}}(g)$ with a strong hybridization, evidenced by the strong bending of the branch away from the cavity frequency \( \omega_0 = E_g\) as the coupling strength \(g\) increases.
Since this branch lies below the particle-hole continuum, it is not broadened by matter absorption and its linewidth is therefore solely set by the cavity linewidth \(\kappa\). 
Consequently the corresponding polariton energy shift $\delta \omega_{\mathrm{P}}$, defined as the difference between the bare photon frequency $\omega_0$ and the polaritonic branch \(\omega_\text{P}\),
must exceed the value of $\kappa$ for the lower polariton to be spectroscopically resolved. This observability condition is assessed quantitatively for a realistic ultracold-atom implementation in \cref{ssec:Feasibility}.
To probe this spectrum, we note that due to the hybridization between matter and light within the cavity, polaritons become the new fundamental modes of the system. Therefore, any frequency-resolved linear-response measurement has access to the polariton energies and thus to their splitting. In the specific context of cavity-QED with ultracold atomic clouds, the state-of-the-art technique has been used in Ref.~\cite{Mottl2012}. It involved Bragg-like spectroscopy to detect the same two polariton-branches from atomic or photonic observables.

We note that, even in the absence of intrinsic subgap excitations, as in the ultracold-atom implementation discussed in \cref{sec:Ultra-imp}, the zero-subgap-absorption picture could in principle be spoiled by a finite fermionic lifetime, which broadens particle-hole excitations.
However, as estimated in \cref{app:subgapAbs}, this effect is too small to appreciably affect the subgap polariton branch for realistic parameters.

\begin{figure}[tbp]
    \centering
\includegraphics[width=\linewidth]{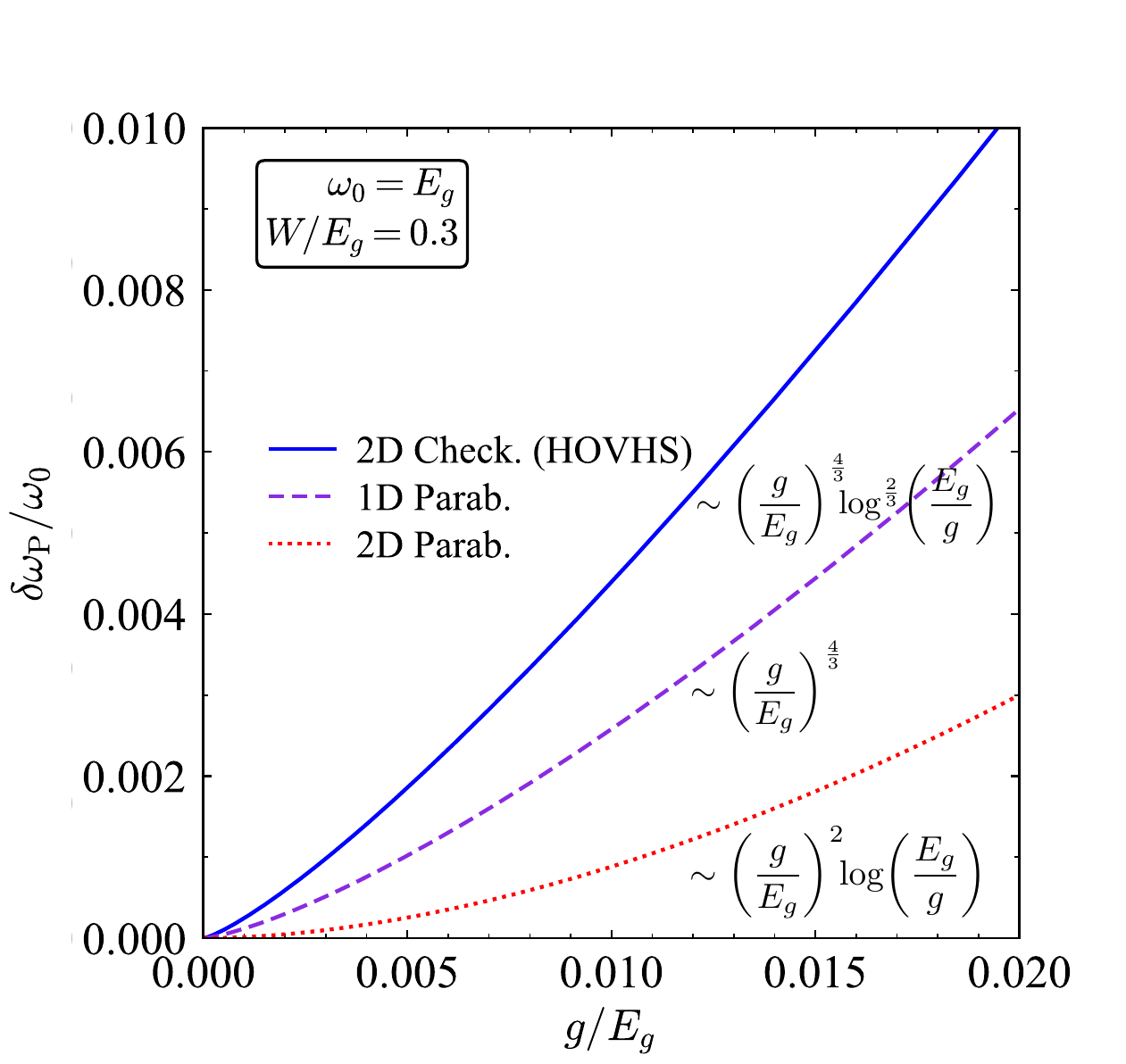}
\caption{The energy shift $\delta\omega_{\mathrm{P}}=\omega_0-\omega_{\mathrm{P}}$ of the lower polariton branch from the bare photon frequency $\omega_0=E_g$ is plotted as a function of the coupling strength $g$ for three cases of insulating band structures: 2D parabolic (red dotted line), 1D parabolic (purple dashed line), and 2D checkerboard (blue  continuous line). The scaling laws as a function of $g/E_g$ are reported next to the curves.
}
    \label{fig:Branch}
\end{figure}

The polariton energy shift $\delta \omega_{\mathrm{P}}$ is directly related to the order of the VHS:
the stronger the divergence at the singularity, the greater the shift.
This is further highlighted in Fig.~\ref{fig:Branch}, where $\delta \omega_{\mathrm{P}}$ is plotted for three different types of insulating band structures: 2D parabolic, 1D parabolic and 2D checkerboard. In the weak light-matter-coupling regime, the different kind of VHS in the three configurations implies a different scaling law of $\delta \omega_{\mathrm{P}}$ with the coupling $g$, written in \cref{fig:Branch} next to the curves. 
As we see, reducing the dimensionality from 2D to 1D increases the energy shift, as proposed in the context of nanowires in Ref.~\cite{Arnardottir2013}.
Notably, however, an even better scaling law is obtained for the 2D checkerboard case, highlighting band engineering of HOVHS in the JDOS as a more effective strategy than dimensionality reduction to enhance polariton hybridization. 

We finally remark that the above scaling results are robust against temperature corrections. In this regard, in \cref{app:finiteT} we show that finite temperature only reduces the strength of the vertical interband response through the thermal population imbalance between valence and conduction bands, encoded in a prefactor \(\mathcal{F}_{T}=\tanh( E_g/4\,k_B T)\): for instance, in the checkerboard case the lower-polariton shift becomes \(\delta\omega_{\mathrm{P,ch}}(T)\sim \mathcal{F}_{T}^{2/3}(g/E_g)^{4/3}\log^{2/3}(g/E_g)\) up to temperature-independent prefactors, so that the weak-coupling scaling shown in \cref{fig:Branch} and later summarized in \cref{tab:sigma_behaviors} is unchanged.

\section{Van Hove Polaritons}\label{sec:Van_Hove_Pol}

\begin{figure}[tbp]
    \centering
\includegraphics[width=\linewidth]{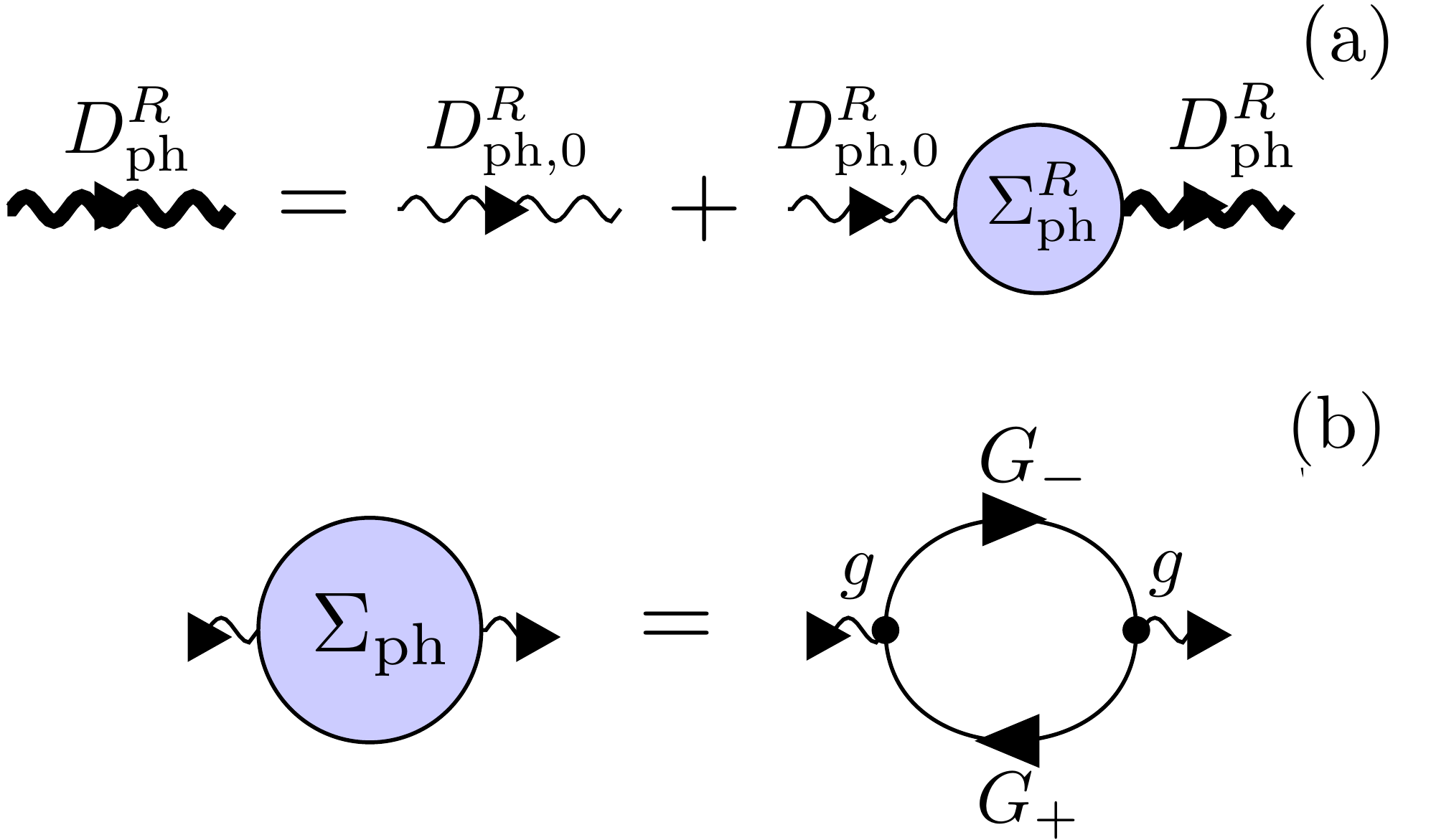}
\caption{ Sketch of (a) the Dyson equation for the photon propagator (wiggly line) and (b) its self-energy due to the interaction with valence ($-$) and conduction ($+$) band electrons.}
    \label{fig:BubbleDiagram}
\end{figure}

Without resorting to a specific lattice model for the matter, in this section we derive the role played by the JDOS of insulating bands on polariton formation, which allows us to discuss the effect of the VHS.  
Employing a single-particle description for the insulating phase, the Hamiltonian in quasi-momentum space can be written as
\begin{align}
\hat{\mathcal{H}} = &{\hat{\mathcal{H}}_{\mathrm{ph}}+\hat{\mathcal{H}}_M+\hat{\mathcal{H}}_{\mathrm{int}} =}\; \omega_0\hspace{0.03cm}\hat{a}^\dagger \hat{a} 
+ \sum_{\lambda = \pm} \sum_{\bm{k} \in \text{BZ}} E_\lambda(\bm{k})\hspace{0.03cm} 
\hat{c}^\dagger_{\lambda, \bm{k}} \hat{c}_{\lambda, \bm{k}} \nonumber \\
& + \sum_{\bm{k} \in \text{BZ}} 
\frac{g(\bm{k})}{\sqrt{N}} \left( 
\hat{a}+\hat{a}^\dagger\right) 
\big(\hat{c}^\dagger_{+, \bm{k}} 
\hat{c}_{-, \bm{k}}+\hat{c}^\dagger_{-, \bm{k}} 
\hat{c}_{+,\bm{k}}\big). 
\label{eqn:Hamiltonian}
\end{align}
The first term represents free photons in a single cavity mode with frequency $\omega_0$, where \(\hat{a}^\dagger\) and \(\hat{a}\) are the photonic creation and annihilation operators. The second term accounts for fermions in the insulating phase, where \(E_\lambda(\bm{k})\) is the energy of the band \(\lambda\) (\(+\) for conduction, \(-\) for valence) at quasi-momentum \(\bm{k}\) belonging to the $1^{\text{st}}\text{ BZ}$, and \(\hat{c}^\dagger_{\lambda, \bm{k}}\) and \(\hat{c}_{\lambda, \bm{k}}\) are the fermionic creation and annihilation operators. The third term captures the light–matter interaction: $g(\bm{k})$ quantifies the bare coupling between photons and vertical (i.e.~with zero-momentum transfer) interband excitations, scaled by the inverse square root of the number of unit cells, $N$. 
Fundamental to the description of cavity polaritons is the
retarded photon propagator
\( D^{R}_{\mathrm{ph}}(t)=
-\,i\,\theta(t)\,
\bigl\langle\,[\hat a(t),\hat a^{\dagger}(0)]\bigr\rangle ,
\)
with $\theta(t)$ the Heaviside step function and
$\hat a(t)=\mathrm e^{\mathrm i\hat{\mathcal H}t}\,\hat a\,
\mathrm e^{-\mathrm i\hat{\mathcal H}t}$ the photon annihilation operator
in the Heisenberg picture. This propagator can be obtained through the Dyson equation
\begin{equation}
    D^{R}_{\mathrm{ph}}(\omega)=
    \Bigl[\bigl( D^{R}_{\mathrm{ph},0}(\omega)\bigr)^{-1} -\Sigma^{R}_{\mathrm{ph}}(\omega)\Bigr]^{-1},
    \label{eq:Dyson_eq}
\end{equation}
where $ D^{R}_{\mathrm{ph},0}(\omega)=
(\omega-\omega_{0}+i \kappa)^{-1}$ is the bare photon propagator including the cavity loss rate $\kappa$, and $\Sigma^{R}_{\mathrm{ph}}(\omega)$ is the retarded photon self-energy. This equation is represented in terms of Feynman diagrams in \cref{fig:BubbleDiagram}(a).
Once this quantity is calculated, the photon spectral function (represented in \cref{fig:densityplot}) follows directly as \(A(\omega) \;=\;
-\frac{1}{\pi}\,
\operatorname{Im}\bigl[{D}^{R}_\mathrm{ph}(\omega)\bigr]\). The lower polariton branch $\omega_\text{P}$ (used in Fig.~\ref{fig:Branch}) is determined by the lower frequency pole of \({D}^{R}_\mathrm{ph}(\omega)\),  which can be obtained as the solution the implicit equation
\(
\omega-\omega_{0}-\operatorname{Re}\bigl[\Sigma^{R}_{\mathrm{ph}}(\omega)\bigr]=0\).

The self-energy can be computed perturbatively in the electron-photon coupling. This corresponds to the particle-hole bubble diagram represented in \cref{fig:BubbleDiagram}(b). The formal derivation of the above equations is detailed in \cref{app:Deriv.Self}, where we obtain an effective action for the photon field by explicitly integrating out the fermionic degrees of freedom within the path-integral Matsubara formalism.
The resulting expression for the self-energy is \footnote{Note that this quantity, if a momentum-independent coupling $g$ is considered, is proportional to the Matsubara frequency representation of the interband polarization bubble \(
\Pi^R(t) = -i \theta(t)  \langle [ \hat{P}(t), \hat{P}^\dagger(0)] \rangle, \quad \text{with } \hat{P}(t) = \sum_k \hat{c}^\dagger_{-,k}(t)\, \hat{c}_{+,k}(t),
\) computed in the absence of light-matter interaction. This follows from the general relation between the dressed photon propagator and the matter polarization function~\cite{PhysRevB.104.235120}.}
\begin{align}
\label{eq:Sigma_Mats}
\Sigma_{\mathrm{ph}}(\omega_{\nu}) =&
\frac{1}{\beta N}\!\sum_{\bm{k},\omega_{n}}g^{2}(\bm{k})
\Bigl[
   G_{+}\!\bigl(\bm{k},\omega_{n}+\omega_{\nu}\bigr)\,
   G_{-}\!\bigl(\bm{k},\omega_{n}\bigr)
 \nonumber\\
 &+ G_{-}\!\bigl(\bm{k},\omega_{n}+\omega_{\nu}\bigr)\,
   G_{+}\!\bigl(\bm{k},\omega_{n}\bigr)
\Bigr],
\end{align}
where \(G_{\lambda}(k,\omega_{n})=[i\omega_n-E_{\lambda}(\bm{k})]^{-1}\) are the bare
conduction/valence electron propagators,
\(\beta\!=\!1/k_{\mathrm{B}}T\) with $T$ being the temperature of the system, and
\(\omega_{n}=(2n+1)\pi T\) and \(\omega_{\nu}=2 \pi \nu T\) (with $n,\nu$ integers) are bosonic and fermionic Matsubara frequencies, respectively.
The retarded self-energy follows via analytic continuation
\(
\Sigma^{R}_{\mathrm{ph}}(\omega) =
\Sigma_{\mathrm{ph}}\!\bigl(i\omega_{\nu}\!\rightarrow\! \omega+i0^{+}\bigr)\). 
The retarded photon self-energy
$\Sigma^{R}_{\mathrm{ph}}(\omega)$ can always be decomposed into an imaginary part, which sets the photon absorption rate (and thus the polariton damping appearing as a width of a peak in the spectral function), and a real part, which shifts the photon energy. In the zero-temperature limit these components evaluate to
\begin{align}
\text{Im}[\Sigma^R_{\mathrm{ph}}(\omega)] &= -\pi g^2 A_{\rm cell} \big[ J(\omega)-J(-\omega)\big], \label{eq:ImSigma} \\
 \text{Re}[\Sigma^R_{\mathrm{ph}}(\omega)]& =-2g^2 A_{\rm cell}
 \, \ {\mathrm{P.V.}} \int_{0}^{\infty} \! \! d\omega' 
\frac{\omega'J(\omega')}{\omega'^{2}-\omega^{2}}, \label{eq:ReSigma}
\end{align}
where \(\mathrm{P.V.}\)~denotes the Cauchy principal value, \(A_{\rm cell}\) is the area of the unit cell, and we set \(g \equiv g(\bm{k}^{\ast})\), with \(\bm{k}^{\ast}\) being the quasi-momentum critical point at which the interband energy splitting \(
\delta E_{\bm{k}} \;=\; E_{+}(\bm{k}) - E_{-}(\bm{k})
\) takes its global minimum $E_g$.  If this minimum is degenerate, we select the point for which the largest number of quasi-momentum derivatives of $\delta E_{\bm{k}}$ vanish. The rationale for this choice will become clear in~\cref{sec:cVHS}.
Furthermore, we introduced the JDOS
\begin{equation}\label{eqn:JDOS_def}
J(\omega)=\frac{1}{S}\sum_{\bm{k}\in\text{BZ}}
\delta\bigl(\omega-\delta E_{\bm{k}}\bigr),
\end{equation}
where $S=NA_{\rm cell}$ is the surface area of the fermionic system. This quantity quantifies the density of states available for vertical interband transitions. 
\cref{eq:ImSigma,eq:ReSigma} explicitly highlight the importance of this quantity in characterizing the polariton. 
As seen in \cref{fig:sketch}(c), $J(\omega)$ is typically vanishing below the band gap (\(\omega<E_g\)) and presents a VHS above it (\(\omega \gtrsim E_g \)). 
When the photon frequency $\omega$ is tuned to a value just below the band gap $E_g$ of the insulator, this automatically implies no absorption, since \(\text{Im}[\Sigma^R_{\mathrm{ph}}(\omega<E_g)] \propto J(\omega<E_g) = 0\). On the other hand, by performing the frequency integral in \cref{eq:ReSigma}, it can be shown that a singularity in $J(\omega)$ for  \(\omega \gtrsim E_g \) leads to a singularity of equal or greater strength in \(\text{Re}[\Sigma^R_{\mathrm{ph}}(\omega)]\) for \(\omega \lesssim E_g \) (an analytical proof is shown in \cref{app:deriv.Re} for the cases considered). 
This singularity in \(\text{Re}[\Sigma^R_{\mathrm{ph}}(\omega)]\) is then directly reflected by the polariton energy, since the latter is determined by the implicit equation $\delta\omega_{\mathrm{P}}=-\text{Re}[\Sigma^R_{\mathrm{ph}}(\omega=\omega_{\mathrm{P}})]$. This gives rise to the scaling laws of $\delta\omega_{\mathrm{P}}(g)$ represented in \cref{fig:Branch} for different cases.

\begin{figure}[tbp]
    \centering
    \includegraphics[width=\linewidth]{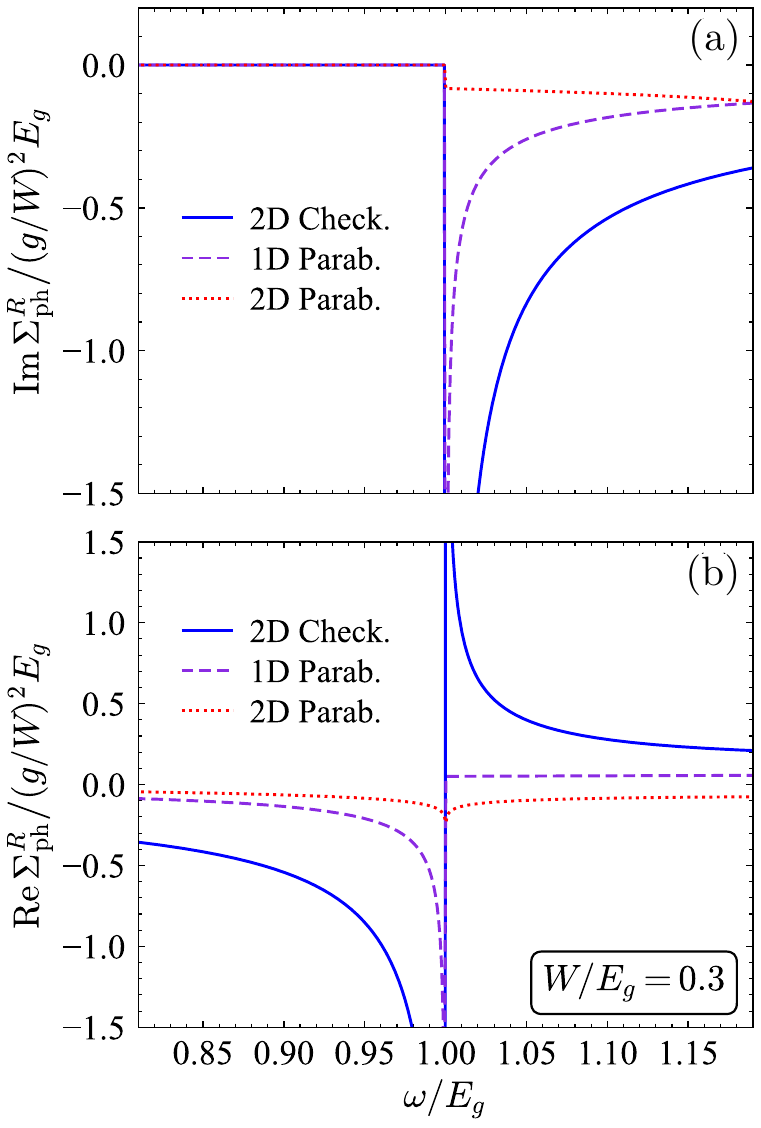}    \caption{ Retarded photonic self-energy $\Sigma_{\mathrm{ph}}^{R}(\omega)$ near the band-gap energy $E_g$ for three model insulators: 2D parabolic (red dotted), 1D parabolic (purple dashed), and 2D checkerboard (blue solid).
(a) Imaginary part $\mathrm{Im}[\Sigma_{\mathrm{ph}}^{R}(\omega)]$.
(b) Real part $\mathrm{Re}[\Sigma_{\mathrm{ph}}^{R}(\omega)]$.}
    \label{fig:Self}
\end{figure}

As already evidenced in \cref{sec:Key_Results}, the strength of the VHS-effect on the self-energy depends on the dimensionality of the system and the local momentum dependence of the interband
transition energy \(
\delta E_{\bm{k}}
\) near \(\bm{k}=\bm{k}^*\). We show this in~\cref{fig:Self}, where we plot the real and imaginary part of $\Sigma^R_{\mathrm{ph}}(\omega)$ for three different types of insulating band structures: 2D parabolic with $\delta E_{\bm{k}}\sim(\bm{k}-\bm{k}^*)^2$, 1D parabolic with $\delta E_{k} \sim(k-k^*)^2$, and 2D checkerboard with  $\delta E_{\bm{k}}\sim[(k_x-k^*_{x})^2-(k_y-k^*_{y})^2]^2$ (see \cref{sec:cVHS} for a derivation of this dispersion within a tight-binding model on a checkerboard lattice). In the 2D parabolic case, the step-like behavior of $\mathrm{Im}[\Sigma^R_{\mathrm{ph}}(\omega)]$ above gap results only in a logarithmic divergence in $\mathrm{Re}[\Sigma^R_{\mathrm{ph}}(\omega)]$. Reducing the dimensionality to 1D, but retaining parabolic bands, results in an enhancement of the above gap divergence of $\mathrm{Im}[\Sigma^R_{\mathrm{ph}}(\omega)]$ to an inverse square root, which corresponds to divergence of equal strength in $\mathrm{Re}[\Sigma^R_{\mathrm{ph}}(\omega)]$. Finally, in the 2D checkerboard case, the engineered nonparabolic bands give rise to a novel type of HOVHS in the JDOS $J(\omega)$, which then appears in \(\text{Im}[\Sigma^\text{R}_{\mathrm{ph}}(\omega)]\) and is reflected into \(\text{Re}[\Sigma^\text{R}_{\mathrm{ph}}(\omega)]\). The singularity features an inverse square-root divergence further enhanced by a logarithmic factor (see \cref{app:cDOSLog,app:deriv.Re} for the derivation):
\begin{align}
   &J(\omega \to E_g^{+}) \sim {\log{\left(\frac{\omega - E_g}{E_g}\right)}}\big{/}{\sqrt{\frac{\omega - E_g}{E_g}}},\label{eqn:Div_Rho}\\
    &\text{Re}[\Sigma^R_{\mathrm{ph}}(\omega \to E_g^{-})] \sim {\log{\left(\frac{E_g - \omega}{E_g}\right)}}\big{/}{\sqrt{\frac{E_g - \omega}{E_g}}},\label{eqn:Div_Re}
\end{align}
which is the strongest divergence of the three considered cases.
The singular behaviors of the real and imaginary parts of the self-energy for the three insulating cases shown in \cref{fig:Self} are summarized in~\cref{tab:sigma_behaviors}, along with the corresponding scaling laws for the polariton energy shift $\delta\omega_{\mathrm{P}}$ shown in \cref{fig:Branch}.
\begin{table}[htb]
\centering
\small
\setlength{\tabcolsep}{1.5pt}      % Slightly reduced column spacing
\renewcommand{\arraystretch}{1.8} % Adjust row height for compactness

\begin{tabular}{|l|c|c|c|}
\hline
\textbf{} & 
$\displaystyle \Im\Sigma_{\mathrm{ph}}^{R}(\omega \!\sim\! E_g^{+})$ & 
$\displaystyle \Re\Sigma_{\mathrm{ph}}^{R}(\omega \!\sim\! E_g^{-})$ & 
$\displaystyle \delta\omega_{\mathrm{P}}$ \\ \hline

\makecell{\textbf{2D} \\ \textbf{par.}} & 
$\text{const.}$ & 
$\log\left(1-\omega/E_g\right)$ & 
$(\frac {g}{E_g})^{2} \log (\frac {g}{E_g})$ \\ \hline

\makecell{\textbf{1D} \\ \textbf{par.}} & 
$\dfrac{1}{\sqrt{\omega/E_g - 1}}$ & 
$\dfrac{1}{\sqrt{1-\omega/E_g}}$ & 
$(\frac {g}{E_g})^{4/3}$ \\ \hline

\makecell{\textbf{2D} \\ \textbf{ch.}}
& 
$-\dfrac{\log(\omega/E_g - 1)}{\sqrt{\omega/E_g - 1}}$ & 
$\dfrac{\log(1-\omega/E_g)}{\sqrt{1-\omega/E_g}}$ & 
$(\frac {g}{E_g})^{4/3}\log^{2/3}(\frac {g}{E_g})$ \\ \hline
\end{tabular}

\caption{Singular behavior of the imaginary and real parts of the retarded photon self-energies in the limits $\omega \to E_g^{\pm}$, and the corresponding polariton energy shifts $\delta\omega_{\mathrm{P}} = \omega_0-\omega_{\mathrm{P}}(g)$. The results are shown for three insulating band structures: 2D parabolic, 1D parabolic, and 2D checkerboard lattice.}
\label{tab:sigma_behaviors}
\end{table}
\section{Engineered HOVHS with Fermions on a Checkerboard Lattice}\label{sec:cVHS}
In this section, we show how a tight-binding model on a 2D checkerboard lattice can be used to engineer an insulating phase with a band dispersion around the gap featuring the desired HOVHS in the JDOS.
\begin{figure}[tb]
    \centering
\includegraphics[width=0.83\linewidth]{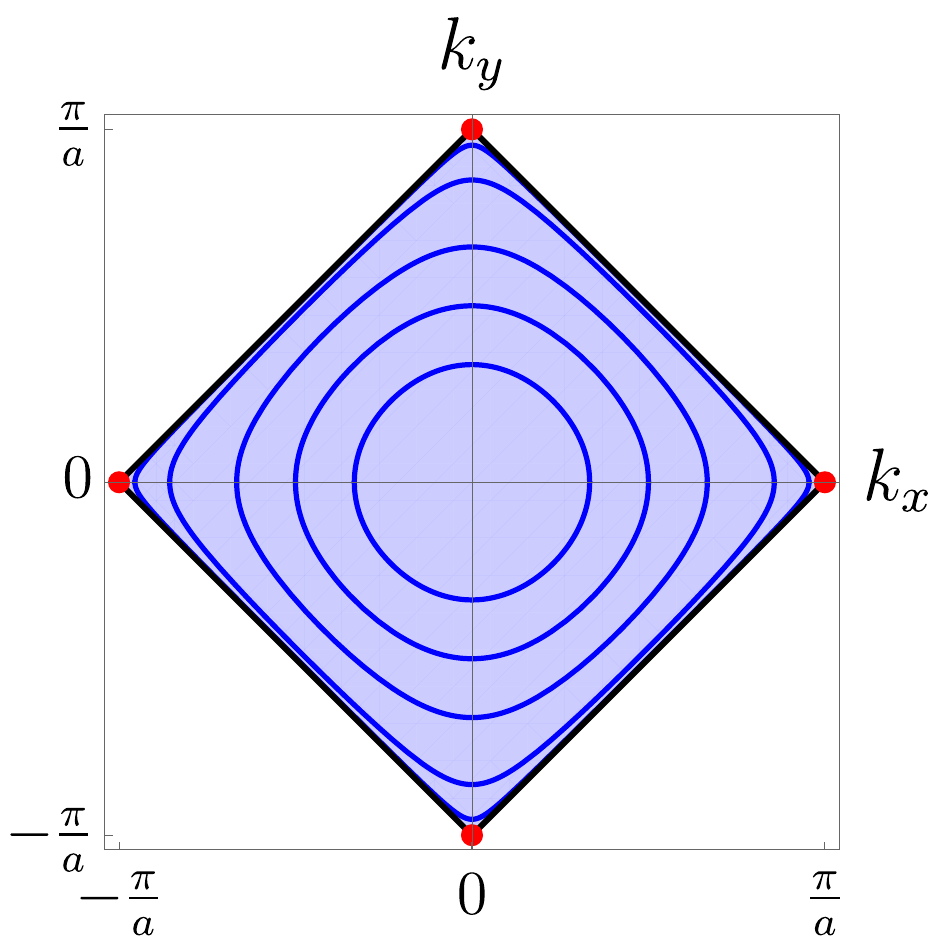}
\caption{Constant-energy contours of the dispersions $E_\pm(\bm{k})$ within the diamond-shaped first BZ of the checkerboard lattice.
Blue lines trace the contours, the black outline marks the BZ boundary, and the red dots indicate the M point and its symmetry-related images at the four vertices.
S}
    \label{fig:BZ_Cheq}
\end{figure}
Let us consider a model of spinless fermions on a square lattice with a nearest-neighbor (NN) hopping of amplitude \(t\) and a staggered on-site potential of amplitude $E_g$. The corresponding Hamiltonian is given by:
\begin{align}
{\hat{\mathcal{H}}_M} = & -t \! \! \sum_{\mathbf{R}_j \in \Lambda_\mathrm{A}}\sum_{l=1}^{4} 
\left( \hat{c}_{\mathrm{A},\mathbf{R}_j}^\dagger \hat{c}_{\mathrm{B},\mathbf{R}_j+\bm{\delta}_l} + \text{h.c.} \right) \nonumber \\
& + \frac{E_g}{2} \left( \sum_{\mathbf{R}_j \in \Lambda_\mathrm{A}} \hat{n}_{\mathrm{A},\mathbf{R}_j} 
- \sum_{\mathbf{R}_j \in \Lambda_\mathrm{B}} \hat{n}_{\mathrm{B},\mathbf{R}_j} \right).
\label{eqn:Matt-Hamil-Tight-NN}
\end{align}
where fermionic creation (annihilation), and number operators \(\hat{c}^{\dagger}_{\alpha,\mathbf{R}_j}\) \((\hat{c}_{\alpha,\mathbf{R}_j}\)), \(\hat{n}_{\alpha,\mathbf{R}_j}\) have been introduced, respectively. These operators refer to fermions located at sites \((\alpha, \mathbf{R}_j)\), where the index \(j\) indicates the \(j\)-th site of the sublattice \(\{\Lambda_\alpha\}\), and the index \(\alpha\) (= A, B) specifies the sublattice. For terms describing hopping between NNs, the positions of the B-type sites are relative to the A-type sites, with the NN vectors \(\{\bm{\delta}_1= (a,0);\bm{\delta}_2=(0,a); \bm{\delta}_3=-\bm{\delta}_1; \bm{\delta}_4=-\bm{\delta}_2\}\), where $a$ is the lattice spacing between A and B sites. This Hamiltonian can be easily diagonalized by performing a Fourier transform 
\(\hat{c}_{\alpha,\mathbf{R}_j} = \frac{1}{\sqrt{N}} \sum_{\bm{k}\in \mathrm{BZ}} e^{-i\bm{k} \cdot \mathbf{R}_j} \hat{c}_{\alpha,\bm{k}}\),
where \( N \) is the number of unit cells and the sum over \(\bm{k}\) runs over the $1^{\text{st}}\text{ BZ}$ shown in Fig.~\ref{fig:BZ_Cheq}, and by applying the canonical basis change
\begin{equation}\label{eqn:Bloch_Basis}
    \begin{split}
   \hat{c}_{-,\bm{k}} &= u_{\bm{k}} \hat{c}_{\mathrm{A},\bm{k}} - v_{\bm{k}} \hat{c}_{\mathrm{B},\bm{k}},\\
   \hat{c}_{+,\bm{k}} &= v_{\bm{k}} \hat{c}_{\mathrm{A},\bm{k}} + u_{\bm{k}} \hat{c}_{\mathrm{B},\bm{k}},
   \end{split}
\end{equation}
where $u_{\bm{k}}=\sin{\theta_{\bm{k}}/2}$,   $v_{\bm{k}}=\cos{\theta_{\bm{k}}/2}$ and their ratio is given by $\tan{\theta_{\bm{k}}}=\varepsilon_{\bm{k}}/E_g$ with \( \varepsilon_{\bm{k}}= -2t(\cos{k_x a}+\cos{k_y a})\). The diagonalized Hamiltonian is then:
\begin{equation}
    \hat{\mathcal{H}}_M= \sum_{\lambda = \pm} \sum_{\bm{k} \in \text{BZ}} E_\lambda(\bm{k})\hspace{0.03cm} 
    \hat{c}^\dagger_{\lambda, \bm{k}} \hat{c}_{\lambda, \bm{k}} \,,
\end{equation}
where the energy bands are given by 
\begin{equation}\label{eqn:bands_tight}
    E_{\pm}(\bm{k})=\pm \sqrt{\varepsilon^2_{\bm{k}}+(E_g/2)^2}.
\end{equation}
The direct gap is obtained when \(\varepsilon_{\bm{k}}=0\), for which
\(E_{+}(\bm{k})=E_g/2\) and \(E_{-}(\bm{k})=-E_g/2\). Since the bands in \cref{eqn:bands_tight} are particle-hole symmetric, the bandwidths of the conduction and valence bands coincide and are given by
\begin{align}
W
&\equiv
\max_{\bm{k}\in\mathrm{BZ}} E_{+}(\bm{k})
-
\min_{\bm{k}\in\mathrm{BZ}} E_{+}(\bm{k})\notag\\
&=
\sqrt{\left({E_g}\big{/}{2}\right)^2+\left(4t\right)^2}
-
{E_g}\big{/}{2},
\label{eqn:checkerboard_bandwidth}
\end{align}
where we used \(\max_{\bm{k}\in\mathrm{BZ}}|\varepsilon_{\bm{k}}|=4t\). Consequently, the vertical interband excitation energy
\(\delta E_{\bm{k}}=E_{+}(\bm{k})-E_{-}(\bm{k})\)
spans the interval \(E_g\leq \delta E_{\bm{k}}\leq E_g+2W\).
The energy bands of \cref{eqn:bands_tight} were represented as solid blue lines in \cref{fig:sketch}(b), and their contour lines within the BZ are represented in solid blue in \cref{fig:BZ_Cheq}.
By Taylor expanding the dispersion near any of the four M points $\bm{k}_\mathrm{M}=\{(\pm\pi/a,0),(0,\pm\pi/a)\}$, highlighted in red in~\cref{fig:BZ_Cheq}, 
we obtain
\begin{equation}\label{eqn:Energy_Bands}
E_\lambda(\bm{k}) \approx \lambda\left\{ \frac{E_g}{2} + \gamma a^4\left[ \left(k_x-k_{\mathrm{M},x}\right)^2 - \left(k_y-k_{\mathrm{M},x}\right)^2\right]^2 \right\}, 
\end{equation}
where $\lambda=\pm$ and we defined $\gamma= { t^2}/{E_g }$ .
The fact that the Hessian matrix \(\partial_{k_i} \partial_{k_j} E_\lambda(\bm{k})\) vanishes identically at the M point makes the latter a critical point associated with a type-I HOVHS~\cite{Yuan2019HighOrderVHS}. The scaling behavior of this singularity in the JDOS was presented in~\cref{eqn:Div_Rho}, and its full analytical derivation is provided in~\cref{app:cDOSLog}. The divergence appears as the product of two contributions at the band gap energy: an inverse square root factor and a logarithmic factor. The origin of this behavior can be understood as follows. The BZ boundary, shown in black in~\cref{fig:BZ_Cheq}, forms a rotated square whose edges are dispersion-flat and have zero group velocity, thereby constituting critical lines. These critical lines are responsible for the one dimensional inverse square root factor, whereas the additional logarithmic factor originates from the intersection of the critical lines at the square’s M vertices. We note that a conventional 2D VHS, which shows a purely logarithmic divergence, arises from a quadratic saddle point situated at the crossing of two noncritical iso-energy lines~\cite{VanHove1953}. 
To the best of our knowledge, the singularity reported here has not been previously documented, because it is associated with a local minimum of the dispersion and has a infinite codimension (see Section VII of~\cite{Class_HOVHS}), while current VHS classifications have thus far addressed only saddle-point scenarios with finite codimension~\cite{Class_HOVHS}.
\section{Implementation: Ultracold Atoms in Optical Lattices Coupled to a Single--Mode Cavity}\label{sec:Ultra-imp}
In this section, we present a physical implementation of the Hamiltonian of \cref{eqn:Matt-Hamil-Tight-NN} with a system of fermionic ultracold atoms in a checkerboard optical lattice. We then propose a cavity-QED setup where the full light-matter Hamiltonian of \cref{eqn:Hamiltonian} can be implemented. 
Finally we provide realistic estimates of the checkerboard parameters in \cref{eqn:Hamiltonian} and assess observability by comparing the expected polariton shift $\delta\omega_{\mathrm P}$ with the cavity linewidth $\kappa$.
\subsection{Matter}\label{ssec:Matter-Model}
\begin{figure}
     \centering     \includegraphics[width=0.95\linewidth]{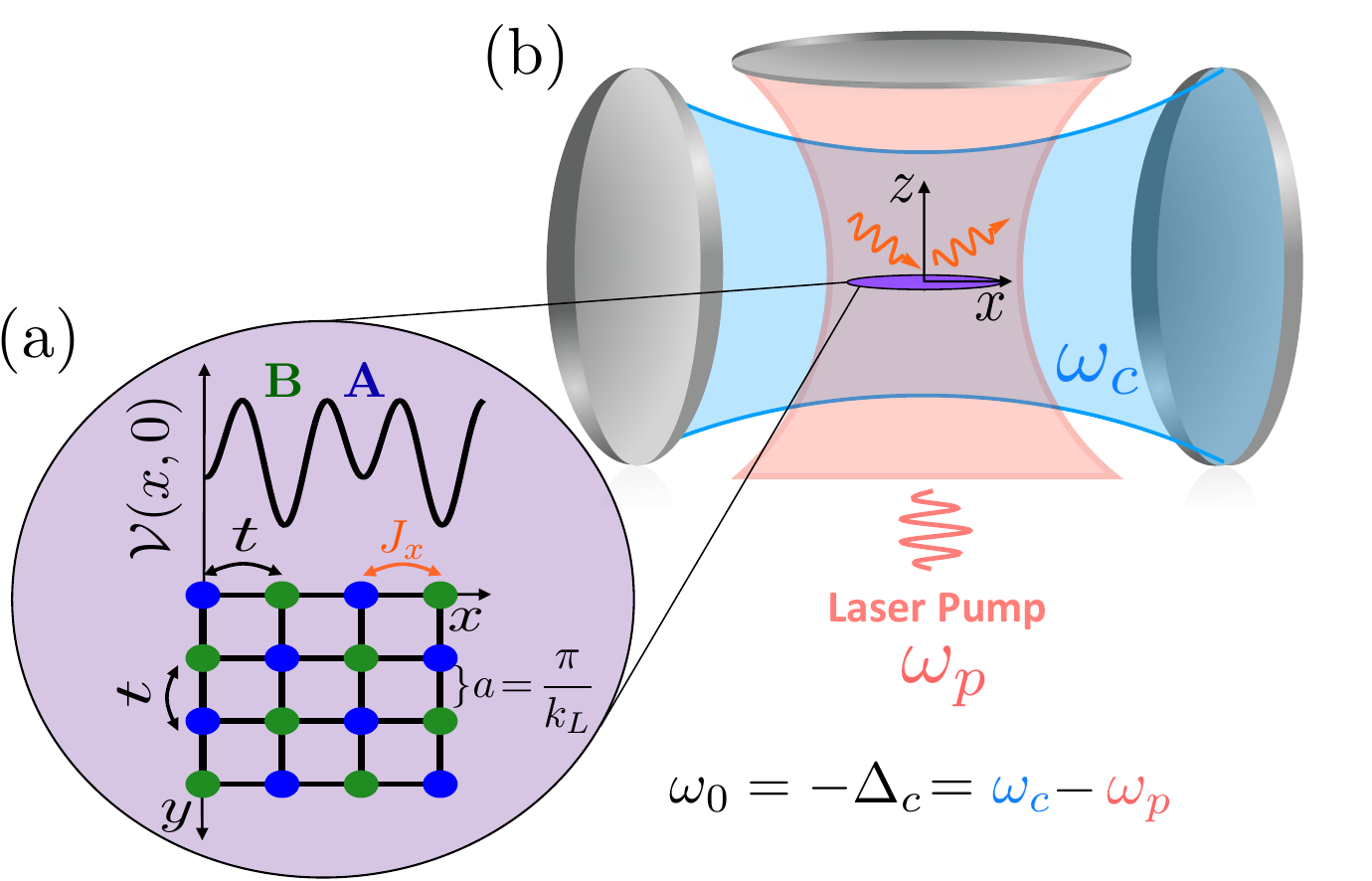}
     \caption{(a) \textit{Bottom}: checkerboard lattice in the $x$-$y$ plane with isotropic hopping $t$ between sublattice sites $A$ and $B$; cavity photons couple to the fermions via cavity-assisted hopping $J_x$ along the $x$ direction.
\textit{Top}: cut at $y=0$ of the optical  potential $\mathcal{V}(\bm{r})$ of \cref{eqn:DW_Potential}.
(b) A 2D Fermi gas, confined by the lattice of panel (a), is loaded into a linear standing-wave cavity of resonance frequency $\omega_c$ whose axis lies along $x$. The atoms are driven by a transverse standing-wave pump laser of frequency $\omega_p$ propagating along $z$. The transverse pump and intracavity lattices are shaded in blue and pink, respectively.
}
     \label{fig:sketch-impl}
 \end{figure}

Let us consider an  ultracold gas of spin-polarized fermionic atoms of mass $m$. 
We assume the cloud to be strongly confined along the $z$ axis, making the system effectively 2D, and trapped in the $x-y$ plane by a periodic potential of the form
\begin{equation}\label{eqn:DW_Potential}
\begin{aligned}
    \mathcal{V}(\bm{r})&= -\mathcal{V}_{0}\hspace{0.03cm}(\cos^2{k_L x}+\cos^2{k_Ly}\\
    &+2\cos{\theta} \, \cos{k_Lx} \,\cos{k_Ly}),
    \end{aligned}
\end{equation}
see Ref.\cite{Wirth2011,Tarruell2012} for possible experimental implementations. This specific potential gives rise to a square optical lattice composed of two classes (A and B) of lattice sites with different well depths arranged in a checkerboard pattern, as shown in Fig.~\ref{fig:sketch-impl}(a). 
The phase $\theta$ in \cref{eqn:DW_Potential} allows to adjust the relative depth of the potential wells at sites A and B.
This system is described by the Hamiltonian $\hat{\mathcal{H}}_M = \int d\bm{r}\hspace{0.03cm} \hat{\psi}^\dagger \hat{H}_M \hspace{0.03cm}\hat{\psi}$, with $\hat{\psi}(\bm{r})$ being the fermionic field operator that annihilates an atom at position $\bm{r}=(x,y)$, and $\hat{H}_M = -(1/2m)\bm{\nabla}^2 +\mathcal{V}(\bm{r}) $ being the single-particle Hamiltonian density. 
Since the gas is spin-polarized, the two-body interactions between fermions are absent, making the single-particle description exact \cite{zwerger2011bcs, zwer14varenna}. We will now derive the Hamiltonian of the checkerboard insulator given in Eq.~\eqref{eqn:Matt-Hamil-Tight-NN}. We assume that the potential wells are deep enough to allow us to map the Hamiltonian $\hat{\mathcal{H}}_M$ onto a tight-binding model on the discrete lattice corresponding to the potential minima. This approximation is implemented by expanding the field operator $\hat{\psi}(\bm{r})$ on a basis \(\{f_{nj}(\bm{r})\}\) of functions localized around each minimum \cite{maschler_ultracold_2008}
\begin{equation}\label{eqn:wannier_exp}
    \hat{\psi}(\bm{r})=\sum_{j} \! \! \sum_{\alpha =A,B} f_{\alpha, j}(\bm{r}) \hat{c}_{\alpha, j} \,,
\end{equation}
where the fermionic operators $\hat{c}_{\alpha, j}$ were defined in~\cref{sec:cVHS}. In the presence of a double well in the unit cell, the functions \(f_{\alpha,j}(\bm{r})\) have to be interpreted as the 2D-version of the generalized Wannier functions proposed by Modugno in Ref.~\cite{Modugno2012}. In the following, we will restrict the analysis to the two lowest energy bands, in analogy with the single-band approximation for the Fermi–Hubbard model~\cite{Ashcroft1976}. Then, within this approximation, the single-particle Hamiltonian can be written as:
\begin{equation}\label{eqn:Matt-Hamil-Tight}
\hat{\mathcal{H}}_M \approx \sum_{\alpha \alpha'= A,B} \sum_{j j'}\bra{f_{\alpha, j}} \hat{H}_M\ket{f_{\alpha', j'}}\hat{c}^{\dagger}_{\alpha, j}\hat{c}_{\alpha', j'}
\end{equation} 
Here the expansion coefficients correspond to the on-site sublattice energies \(\epsilon_\alpha = \bra{f_{\alpha, j}} \hat{H}_M\ket{f_{\alpha, j}}\), and to minus the tunneling amplitudes between different subwells \(t^{jj'}_{\alpha \alpha'} = -\bra{f_{\alpha, j}} \hat{H}_M\ket{f_{\alpha', j'}}\). Under the tight-binding approximation~\cite{Ashcroft1976} and absorbing into the chemical potential $\mu$ all the elements proportional to the identity, the Hamiltonian in Eq.~\eqref{eqn:Matt-Hamil-Tight} can be mapped to the Hamiltonian in Eq.~\eqref{eqn:Matt-Hamil-Tight-NN} with $E_g={\epsilon_A-\epsilon_B}$.

\subsection{Coupling to light}\label{ssec:Light-Matter-Model}
Let us now assume that the atomic cloud trapped in the optical lattice is loaded inside a single-mode linear standing-wave cavity with cavity axis aligned along $x$ and resonance frequency
$\omega_c$.  The atoms are driven by a transverse standing-wave pump of frequency $\omega_p$ with a detuning from the cavity frequency $\Delta_c=\omega_p-\omega_c<0$. The pump is oriented along the $z$ direction, i.e.~orthogonally to both the cavity axis and to the plane in which the atoms are trapped, as sketched in~\cref{fig:sketch-impl}(b).
We operate in the dispersive regime, where the detuning $\Delta_a = \omega_p - \omega_a$ of the pump frequency with respect to the frequency $\omega_a$ of an internal atomic transition is much larger than any other energy scale in the system. Under this condition, the atomic excited state can be adiabatically eliminated, and the dynamics of the atom–cavity system is governed by the effective Hamiltonian~\cite{Mivehvar2021}
\begin{equation}\label{eqn:Hamil}
\begin{aligned}
\hat{\mathcal{H}}
  = &-\Delta_c\,\hat{a}^\dagger\hat{a}\\
  &+\int d\bm{r}\;\hat{\psi}^\dagger(\bm{r})
     \Bigl[
     -\tfrac{\nabla^2}{2m}
     + \mathcal{V}(\bm{r}) \\
  &
     + U(\bm{r})\,\hat{a}^\dagger\hat{a}
     + \eta_{1\text{D}}(\bm{r})\bigl(\hat{a}^\dagger+\hat{a}\bigr)
     \Bigr]
     \hat{\psi}(\bm{r}),
\end{aligned}
\end{equation}
which is written directly in the $z=0$ plane where the atoms are confined. $\mathcal{V}(\bm{r})$ is the 2D checkerboard potential of Eq.~\eqref{eqn:DW_Potential}. The cavity mode generates an intracavity lattice $U(\bm r)=U_{0}\cos^{2}(q_{c}x)$.
The transverse pump creates a standing wave $V(\bm r,z)=V_{0}\cos^{2}(q_{p}z)$ that, at $z=0$, reduces to the uniform offset $V_{0}$; this constant simply renormalizes the chemical potential and is therefore omitted from the Hamiltonian. The interference between pump and cavity fields produces the position-dependent atom–photon coupling
$\eta(\bm r,z)=\eta_{0}\cos(q_{c}x)\cos(q_{p}z)$,
which in the plane $z=0$ becomes the one-dimensional periodic coupling $\eta_{1\mathrm D}(\bm r)=\eta_{0}\cos(q_{c}x)$ that enters the Hamiltonian.
The lattice and coupling strengths $U_0$, $V_0$ and $\eta_0$ follow from the position-dependent pump Rabi frequency $\Omega(\bm{r},z)=\Omega_0\cos(q_p z)$ and the cavity–matter coupling $\mathcal G(\bm{r})=\mathcal G_0\cos(q_c x)$: $V_0=\Omega_0^{2}/\Delta_a$, $U_0=\mathcal G_0^{2}/\Delta_a$, and $\eta_0=\Omega_0\mathcal G_0/\Delta_a$~\cite{Mivehvar2021}. Additionally we choose $q_c=2k_L$, so that the periodicity of $\eta(\bm{r})$ along $x$ is twice the one of $\mathcal{V}(\bm{r})$.

To show that this system hosts HOVHS-polaritons, it is sufficient to map the Hamiltonian~\eqref{eqn:Hamil} onto the effective Hamiltonian~\eqref{eqn:Hamiltonian}. To this end, 
the matter sector has already been mapped in Sec.~\ref{ssec:Matter-Model}. Mapping the photonic part is straightforward once we identify the bare cavity frequency $\omega_0$ with the negative detuning $-\Delta_c > 0$. 
We remark that this mapping allows to address the lattice energy gap $E_g$, which is typically of the order of a few recoil energies, by employing cavity and pump frequencies $\omega_c$ and $\omega_p$ in the optical regime, i.e.~at much larger energies. The reason is that the effective Hamiltonian \eqref{eqn:Hamil} is derived in the frame of the pump rotating at frequency $\omega_p$, so that the energy matching happens at the level of the energy difference  $\omega_0\equiv-\Delta_c=\omega_c-\omega_p$, not at the level of the absolute frequencies $\omega_c$ and $\omega_p$. This difference can then be easily tuned to the value $E_g$. In the lab frame, this corresponds to address the gap $E_g$ with a two-photon process involving a pump and a cavity photon.
The residual diamagnetic  contribution $\hat{a}^\dag \hat{a}\int d\bm{r}\, \hat{\psi}^\dag(\bm{r})\hat{\psi}(\bm{r}) U(\bm{r})$ merely produces a uniform dispersive shift and can be absorbed into a redefinition of the cavity detuning $\omega_0\equiv-\Delta_c$, as shown in Ref.~\cite{Piazza2013BEC_DHL}. The only term that still calls for an explicit treatment is the interaction vertex 
\mbox{$(\hat a+\hat a^\dagger)\int d\bm{r}\,\eta_{1\text D}(\bm{r})\,
\hat\psi^\dagger(\bm{r})\hat\psi(\bm{r})$},
which needs to be mapped to $\hat{\mathcal H}_{\mathrm{int}}$ in~\cref{eqn:Hamiltonian}.
The choice \( q_c = 2k{_L} \) makes the periodic function \( \eta_{1\text{D}}(\bm{r}) \) commensurate with the bipartite lattice, thereby allowing us to express $\hat{\mathcal{H}}_{\mathrm{int}}$ in the basis \(\{f_{nj}(\bm{r})\}\) of generalized Wannier functions [see \cref{eqn:wannier_exp}]:
$\hat{\mathcal{H}}_{\mathrm{int}} = \left(\hat{a}+\hat{a}^\dagger\right) \sum_{\alpha \alpha'= A,B} \sum_{j j'}\bra{f_{\alpha, j}} \hat{\eta}_{1\text{D}}\ket{f_{\alpha', j'}}\hat{c}^{\dagger}_{\alpha, j}\hat{c}_{\alpha', j'}.$ Neglecting hopping beyond NNs, $\hat{\mathcal{H}}_{\mathrm{int}}$ can be further written as
\begin{align}\label{eqn:Int_Ham}
    \hat{\mathcal{H}}_{\mathrm{int}} & =  \left(\hat{a} + \hat{a}^\dagger\right) 
    \sum_{l=1,3} J_x \sum_{\bm{R}_j \in \Lambda_A} \big( 
\hat{c}_{A,\bm{R}_j}^\dagger \hat{c}_{B,\bm{R}_j + \bm{\delta}_l} 
    + \text{h.c.} \big) \notag \\
    & +  \left(\hat{a} + \hat{a}^\dagger\right) 
    \sum_{l=2,4} J_y \sum_{\bm{R}_j \in \Lambda_A} \big( 
\hat{c}_{A,\bm{R}_j}^\dagger \hat{c}_{B,\bm{R}_j + \bm{\delta}_l} 
    + \text{h.c.} \big) \notag \\
    & + \left(\hat{a} + \hat{a}^\dagger\right)\!\!\sum_{\alpha=\text{A},\text{B}} \!\!C_\alpha \sum_{\bm{R}_j \in \Lambda_\alpha} 
    \hat{n}_{\alpha, \bm{R}_j},
\end{align}
where we defined the sublattice-dependent density-coupling constants $C_\alpha\equiv\bra{f_{\alpha, j}} \hat{\eta}_{1\text{D}}\ket{f_{\alpha, j}}$ and the cavity-assisted hopping amplitudes $J_x\equiv \bra{f_{A,\bm{R}_j}} \hat{\eta}_{1\text{D}}\ket{f_{B, \bm{R}_j+\bm{\delta}_1}} = \bra{f_{A, \bm{R}_j}} \hat{\eta}_{1\text{D}}\ket{f_{B, \bm{R}_j +\bm{\delta}_3}}$
and $J_y\equiv \bra{f_{A, \bm{R}_j}} \hat{\eta}_{1\text{D}}\ket{f_{B, \bm{R}_j+\bm{\delta}_2}}=\bra{f_{A, \bm{R}_j}} \hat{\eta}_{1\text{D}}\ket{f_{B, \bm{R}_j+\bm{\delta}_4}}$.
Note that the equality of the cavity-assisted hopping amplitudes for the two pairs of displacement vectors oriented along the same axis, \((\bm{\delta}_1,\bm{\delta}_3)\) along \(x\) and \((\bm{\delta}_2,\bm{\delta}_4)\) along \(y\), stems from the fact that \(\eta_{1\mathrm D}(\bm r_\parallel)\) is invariant under reflections  centered on the lattice sites. 
In the Bloch basis~\eqref{eqn:Bloch_Basis}, the Hamiltonian $\hat{\mathcal{H}}_{\mathrm{int}}$ takes the form \begin{equation}\label{eqn:Int-Hamil-Bloch}
\hat{\mathcal{H}}_{\mathrm{int}} = \left(\hat{a}+\hat{a}^\dagger\right) \sum_{\lambda}\sum_{\bm{k}\in \mathrm{RBZ}}\frac{g_{\lambda\lambda'}(\bm{k})}{\sqrt{N}}
\hat{c}^\dagger_{\lambda,k}\hat{c}_{\lambda',k},
\end{equation}
where we defined the intraband couplings $g_{\lambda\lambda}(\bm{k})\big{/}\sqrt{N}=\bar{C} + \lambda\left(\frac{\delta C E_g/2+ 2  \varepsilon_{\bm{k}}\xi_{\bm{k}}}{\sqrt{E_g^2+4\varepsilon^2_{\bm{k}}}}\right)$ and the interband couplings $g_{+-}(\bm{k})\big{/}\sqrt{N}=g_{-+}(\bm{k})\big{/}\sqrt{N}= \left(\frac{\delta C \varepsilon_{\bm{k}} - \xi_{\bm{k}}E_g}{\sqrt{E_g^2+4\varepsilon^2_{\bm{k}}}}\right)$ with $\bar{C}= \frac{C_A + C_B}{2}$, 
$\delta C= C_A - C_B$ and $\xi_{\bm{k}}=-2J_x\cos{k_x a} -2J_y\cos{k_y a}$. We recall from~\cref{sec:cVHS} that $\varepsilon_{\bm{k}}=-2t(\cos{k_x a}+\cos{k_y a})$. The mapping is complete once the interband coupling is identified as \( g(\bm{k}) = g_{+-}(\bm{k}) \).
Intraband terms can be neglected as they do not contribute to the photonic self-energy due to the absence of quasi-momentum transfer in the interaction. This anisotropy is crucial: if the hopping were isotropic, i.e.~$J_x = J_y$, the interband coupling would vanish at $\bm{k}^*$.

\subsection{Estimate of realistic experimental parameters}\label{ssec:Feasibility}
Our predictions are general and do not rely on a specific atomic species. Here, to connect with current experimental capabilities, we focus on the widely used $^{6}$Li platform~\cite{Zhang2021ScienceSuperradFermi,Helson2023NatureDW,Buhler2025SciPostCODT,Zwettler2025b}.
We take a cavity wavevector $q_c$ such that the corresponding recoil energy is $E_{r,c}=\hbar^{2}q_c^{2}/2m=h\times 73.67\,{\rm kHz}$ \cite{Helson2023NatureDW}.
In the geometry discussed in \cref{ssec:Light-Matter-Model} we choose set $q_c=2k_L$, so that it is convenient to introduce also the recoil energy associated with the checkerboard lattice, $E_{r,L}=\hbar^{2}k_L^{2}/2m=E_{r,c}/4$.
To estimate the tight-binding parameters entering \cref{eqn:Hamiltonian} and the cavity-assisted hopping amplitudes appearing in \cref{eqn:Int_Ham}, we compute maximally localized Wannier functions for the checkerboard potential \cref{eqn:DW_Potential} using the \texttt{Wannier90} code \cite{Pizzi2020Wannier90}.
For $\mathcal{V}_0=5E_{r,L}$ and a $\theta\simeq0.424\pi$ we estimate, from the maximally localized Wannier functions, the band gap
$E_g\simeq 3.18\,E_{r,L}$, the NN hopping
$t\simeq0.17\,E_{r,L}$, and the cavity-assisted hopping matrix elements
\begin{align}
J_x &= \big\langle f_{A,\bm R_j}\big|\cos(2k_L x)\big|f_{B,\bm R_j+\bm\delta_1}\big\rangle \,\eta_0
\simeq (0.0270)\,\eta_0,\\
J_y &= \big\langle f_{A,\bm R_j}\big|\cos(2k_L x)\big|f_{B,\bm R_j+\bm\delta_2}\big\rangle \,\eta_0
\simeq (-0.0086)\,\eta_0,
\end{align}
where $\eta_0$ is defined in \cref{ssec:Light-Matter-Model}.
In the checkerboard model, the bandwidth parameter $W$ is related to $t$ and $E_g$ by
$
{W}=\sqrt{(E_g/2)^2+(4t)^2}-E_g/2.
$
With the above Wannier estimates $t/E_g\simeq0.053$, we find $W/E_g\simeq0.044$.
This value is smaller than the representative ratio used in some of the illustrative plots elsewhere in the manuscript (e.g.\ $W/E_g=0.3$), implying quantitative changes in prefactors, but leaving the scaling laws in \cref{tab:sigma_behaviors} unchanged.
Recent experiments report maximum values of the cavity potential depth per photon and pump light shift corresponding to $U_0/h\simeq 20\,{\rm Hz}$ and $V_0/h\simeq 10\, E_{r,c}\simeq40\,\text{kHz}$  \cite{Helson2023NatureDW,Buhler2025SciPostCODT,Zwettler2025b}.
Using $\eta_0=\sqrt{U_0V_0}$, this gives
$
\eta_0\simeq 0.052\,E_{r,c},
$
where we used $E_{r,c}/h=73.67\,{\rm kHz}$ \cite{Helson2023NatureDW}.
As discussed in \cref{ssec:Light-Matter-Model}, the relevant interband coupling at the checkerboard HOVHS is controlled by the anisotropy $\delta J=J_x-J_y$, which is nonzero for $J_x\neq J_y$. The number of unit cells \(N\) participating in the light--matter coupling can be estimated from the transverse size of the atomic cloud.
We use the harmonic-trap characterization of Ref.~\cite{Helson2023NatureDW}, where \(N_{\rm at}=3.5\times 10^5\) and the geometric mean of the trap frequencies in the hybrid trap is
\(\bar{\omega}=(\omega_x\omega_y\omega_z)^{1/3}=2\pi\times 106\,{\rm Hz}\).
The corresponding Fermi energy is defined by
\( E_{\mathrm{Fh}}=\hbar \bar{\omega}(3N_{\rm at})^{1/3}
\). The associated Thomas--Fermi radius is
\(
R_{\mathrm{TF}}=({2E_{\mathrm{Fh}}}/{m_{\rm Li}\bar{\omega}^2})^{1/2}\simeq 56.8\,\mu{\rm m},
\)
with \(m_{\rm Li}\) the mass of \(^{6}{\rm Li}\).
For the checkerboard geometry considered here, \(q_c=2k_L\), and the lattice spacing is \(a=\pi/k_L\simeq 671\,{\rm nm}\).
Since the checkerboard primitive cell contains two sites, its area is
\(
A_{\rm cell}=2a^2\simeq 0.90\,\mu{\rm m}^2.
\)
Approximating the illuminated two-dimensional cloud as a disk of radius \(R_{\mathrm{TF}}\), we obtain 
\(
 N\simeq {\pi R_{\mathrm{TF}}^2}/{A_{\rm cell}}\simeq 1.1\times 10^4
\) as an estimate of the number of cells.
This is the value entering the \(1/\sqrt{N}\) normalization of the light--matter coupling. As a representative value for cavity decay rate through the mirrors, we  consider \(\kappa/h=4.5\,{\rm kHz}\) from the setup of  Ref.~\cite{Kessler2014NJPSubrecoil}. Using these values in the checkerboard polariton-shift calculation presented in \cref{sec:Van_Hove_Pol}, we find that $\delta\omega_{\mathrm P}\gtrsim \kappa$ already for $\eta_0 \gtrsim 0.009\,E_{r,c}$, compatible with the experimentally achievable value $
\eta_0\simeq 0.052\,E_{r,c},
$ obtained above.
This indicates that the observability condition $\delta\omega_{\mathrm P} \geq \kappa$ can be satisfied in the checkerboard implementation with current experimental capabilities.
\section{Conclusions}
We have shown that band engineering of fermionic insulators coupled to cavity photons allows to enhance hybridization between light and matter via high-order Van-Hove singularities. By tuning the cavity resonance frequency just below the insulating gap, this leads to the formation of polaritons experiencing a strong energy shift without appreciable absorption. This approach provides an alternative to dimensionality reduction proposed in \cite{Arnardottir2013}, that can lead to a better scaling law of the energy shift with the light-matter coupling.

In this work, we have focused on an ultracold atom implementation which is available in state-of-the-are setups. The main reason for this is the absence of lattice excitation, impurities, and the possibility of tuning the atom-atom interactions to zero thereby removing further in-gap modes (like excitons) that would spoil the quality of the Van Hove singularity and its positive effect on the polariton properties. We have also provided a detailed estimate of realistic model parameters, placing our proposal within reach of state-of-the-art experimental implementations.

The prospects for implementing our proposal in solid-state materials are strengthened by the recent observation of subgap polaritons in insulating bilayer graphene coupled to a subwavelength Terahertz resonator~\cite{Helmrich2026}. No intrinsic excitonic resonances were detected in that experiment: an absence attributed to the small exciton binding energies and weak oscillator strengths at the small gaps explored, together with strong Coulomb screening by the nearby gates. Indeed, engineering the surrounding dielectric environment provides a practical means to control this screening and suppress excitonic effects~\cite{raja_coulomb_2017,tebbe_tailoring_2023}. Together with ongoing progress in identifying and engineering HOVHSs, especially in layered materials~\cite{Yuan2019HighOrderVHS,seiler_quantum_2022,Kagome_2022,chandrasekaran_engineering_2024}, this strategy offers a promising route to realizing high-order Van Hove polaritons in solid-state materials.

In this spirit, we would like to conclude by stressing a 
generic advantage of the polaritons discussed here, resulting from hybridization between photons and a continuum of interband particle-hole excitations, with respect to the well established exciton-polaritons \cite{carusotto_RMP,Delteil2019,MunozMatutano2019}. This advantage is relevant for the purpose of (quantum) nonlinear optics and stems from the fact that, while exciton-polaritons inherit their nonlinearity from the residual exciton-exciton interactions (typically weak), the coupling between particle-hole excitations and photons is intrinsically nonlinear: polaritons in this case will interact via the high-order polarization functions of the matter. 
The latter will also be enhanced by Van Hove singularities, as is the lowest order polarization function discussed here as the photon self-energy. This aspect will be the object of future investigation.

\begin{acknowledgments}
The authors thank D.~Efremov, F.~Helmrich, T.~Nova, and T.~Weitz for insightful discussions. The data that support the findings of this article are openly available \cite{GitHubRep}.
\end{acknowledgments}

\appendix
\section*{Summary of the appendices}
In \cref{app:Deriv.Self}, we derive the retarded photon self-energy using a path-integral formulation. 
\Cref{app:cDOSLog} is devoted to the analytical computation of the checkerboard-lattice JDOS. 
In \cref{app:deriv.Re}, we derive the subgap behavior of the real part of the retarded self-energy and in 
\cref{app:deriv.branche} the weak-coupling scaling laws of the lower-polariton energy shift. 
Finally, \cref{app:finiteT} and \cref{app:subgapAbs} studies the potential effects on the polariton shift of finite temperature and subgap absorption due to atomic dissipation, respectively.

\section{Derivation of \(\Sigma^R_{\mathrm{ph}}(\omega)\)}\label{app:Deriv.Self}
\begin{figure}[tbh]
    \centering
\includegraphics[width=0.43\textwidth]{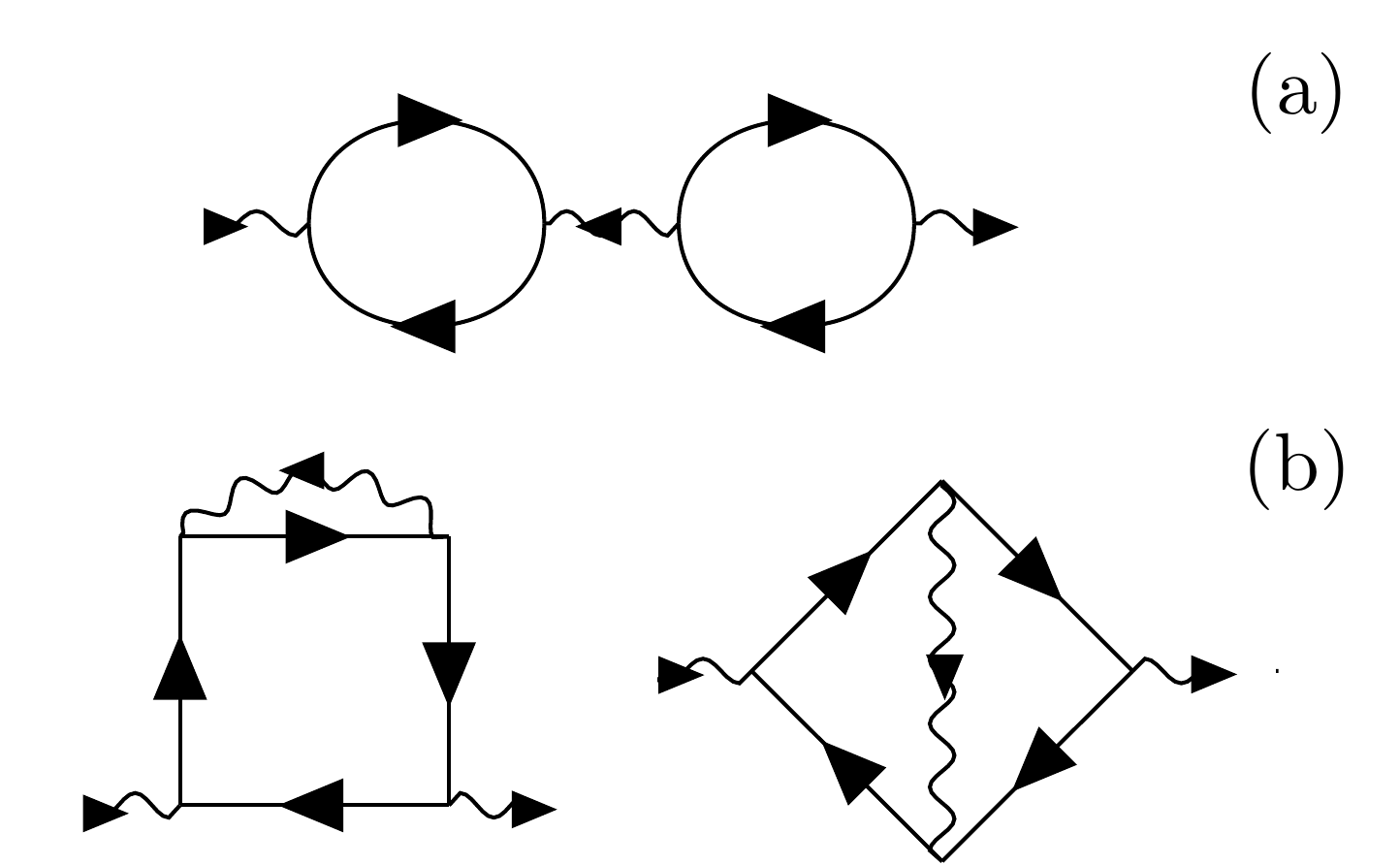}
\caption{Neglected corrections to the polaritonic dispersion and spectral width at fourth order in the coupling $g$. (a) Contribution of off-diagonal terms in \cref{eqn:Nambu_Dinv}.  
(b) Corrections from the effective four-photon vertex interaction.}
    \label{fig:neglected_corrections}
\end{figure}

The aim of this section is to derive the photonic retarded self-energy \(\Sigma^R_{\mathrm{ph}}(\omega)\), whose imaginary and real parts are given in \cref{eq:ImSigma,eq:ReSigma}. This quantity arises naturally within the construction of the photonic effective theory through the path integral formalism on the imaginary time-axis $\tau \in [0,\beta]$, with $\beta$ the inverse temperature.
The action corresponding to the Hamiltonian in Eq.~\eqref{eqn:Hamiltonian} is quadratic in the fermionic fields, which can thus be integrated out exactly. The resulting effective action for the photons, decomposing the field into Matsubara components $\alpha(\tau)=\frac{1}{\beta}\sum_{\omega_\nu}\alpha(\omega_\nu)e^{-i\omega_\nu \tau}$, becomes~\cite{Piazza2013BEC_DHL}
\begin{align}
\mathcal{S}_{\mathrm{ph}}[\alpha,\alpha^*]& =\frac{1}{\beta}\sum_{\omega_\nu} (-i\omega_\nu + \omega_0)\alpha^*(\omega_\nu)\alpha(\omega_\nu)\notag\\
&-\Tr{\log{\mathcal{M[\alpha,\alpha^*]}}},
\end{align}
where we defined the matrix 
\begin{equation}
    \mathcal{M}_{ab}[\alpha,\alpha^*]=-\mathcal{G}_{ab}^{-1}+\mathcal{A}_{ab}[\alpha,\alpha^*],
\end{equation}
where the latin letters denote the full set of fermionic indices $({\bm{k},\omega_n, \lambda})$, and the matrices $\mathcal{G}_{ab}$ and $\mathcal{A}_{ab}[\alpha,\alpha^*]$ are defined as
\begin{align}
\mathcal{G}_{ab}^{-1}&=G^{-1}_\lambda(\bm{k},\omega_n)\delta_{\lambda,\lambda'}\delta_{\omega_n,\omega_n'}\delta_{\bm{k},\bm{k}'}\\
\mathcal{A}_{ab}[\alpha,\alpha^*]&=\frac{g(\bm{k})}{\sqrt{\beta N}}(1-\delta_{\lambda,\lambda'})\delta_{\omega_\nu,\omega_n'-\omega_n} \delta_{\bm{k},\bm{k}'}\notag\\
& \quad \times \left(\alpha(\omega_\nu)+\alpha^*(-\omega_\nu)\right),
\end{align}
with $G_{\lambda}(\bm{k},\omega_n)=[i\omega_n-E_\lambda(\bm{k})]^{-1}$ being the fermionic propagator and the chemical potential $\mu$ was set to zero. In order to compute the photon self-energy used in the main text, we expand the logarithm up to second order
$\Tr{\log{\mathcal M}}\approx\Tr{\log{\mathcal G^{-1}}}+\Tr{\mathcal G \mathcal A}-\frac{1}{2}\Tr{(\mathcal G \mathcal A)^2}$ to obtain the following effective action:
\begin{equation}
\mathcal{S}_{\mathrm{ph}}[\alpha,\alpha^*]=-\frac{1}{2\beta}\sum_{\omega_\nu}\begin{pmatrix} \alpha^*(\omega_\nu)\ \alpha(\omega_\nu) \end{pmatrix}\mathcal{D}^{-1}_{\mathrm{ph}}(\omega_\nu) \! \begin{pmatrix} \alpha(\omega_\nu) \\ \alpha^*(\omega_\nu) \end{pmatrix},
\end{equation}
where the inverse cavity propagator reads:
\begin{equation}\label{eqn:Nambu_Dinv}
    \mathcal{D}^{-1}_{\mathrm{ph}}(\omega_\nu) = \begin{pmatrix} 
i\omega_\nu - \omega_0 -\Sigma_{\mathrm{ph}}(\omega_\nu) & -\Sigma_{\mathrm{ph}}(\omega_\nu) \\ 
-\Sigma_{\mathrm{ph}}(\omega_\nu) & -i\omega_\nu - \omega_0 -\Sigma_{\mathrm{ph}}(\omega_\nu) 
\end{pmatrix},
\end{equation}
with the self-energy given by~\cref{eq:Sigma_Mats}. In our analysis we focus on the diagonal elements [see~\cref{eq:Dyson_eq}],  which describe number-conserving photon processes.
The neglect of the off-diagonal components is justified later in this appendix. Performing the summation on the bosonic Matsubara frequencies and the analytic continuation $i\omega_\nu\rightarrow \omega + i 0^+$, one obtains the retarded photonic self-energy 
\begin{align}\label{eqn:Self-Energy}
\Sigma^R_{\mathrm{ph}}(\omega)&=\frac{1}{N}\sum_{\bm{k}}g^2(\bm{k})\hspace{0.03cm}\Bigg(\frac{n_F\left(E_-(\bm{k})\right)-n_F\left(E_+(\bm{k})\right)}{\omega-(E_+(\bm{k})-E_-(\bm{k}))+i0^{+}}\notag\\
& \quad +\frac{n_F\left(E_+(\bm{k})\right)-n_F\left(E_-(\bm{k})\right)}{\omega+(E_+(\bm{k})-E_-(\bm{k}))+i0^{+}}\Bigg),
\end{align}
with $n_F(E)=1/(e^{\beta E}+1)$ being the Fermi-Dirac distribution. 
At zero temperature the Fermi factors become step functions,
$n_F(E_{\pm})=\Theta(\mu-E_{\pm})$.
Pinning the chemical potential at mid-gap ($\mu=0$) ensures
$E_{-}(\bm{k})<\mu<E_{+}(\bm{k})$ for every $\bm{k}$; hence
$\Theta(\mu-E_{-})=1$ and $\Theta(\mu-E_{+})=0$.
Consequently, the numerators in~\cref{eqn:Self-Energy} collapse to the constants
$\pm[n_F(E_{-})-n_F(E_{+})]=\pm1$. 
 Thus the imaginary part of the retarded photonic self-energy \({\Sigma}_{\mathrm{ph}}^R\) at zero temperature is given by:
\begin{align}\label{eqn:ImSigmadef}
    \Im[\Sigma^{R}_{\mathrm{ph}}(\omega)]&=- \frac{\pi}{N} \!\!\sum_{\bm{k}\in BZ}\!\! g^2({\bm{k}})\Big[\delta\left(\omega-\delta E_{\bm{k}}\right) -\delta\left(\omega+\delta E_{\bm{k}}\right)\Big],
\end{align}
having also employed the identity $
\lim_{\epsilon \to 0} \frac{1}{x + i\epsilon} = \mathrm{P.V.}\left(\frac{1}{x}\right) - i\pi \delta(x)
$ to isolate the imaginary part. The real and imaginary parts satisfy the Kramers--Kronig identities~\cite{Fetter&Walecka}:
\begin{equation}
    \Re[\Sigma^R_{\mathrm{ph}}(\omega)]=
\frac{2}{\pi} \,\mathrm{P.V.} \int_{0}^{\infty}
\frac{\omega'\Im[\Sigma^{R}(\omega')]}{\omega'^{2}-\omega^{2}}\,d\omega'.
\label{eq:ReSigma}
\end{equation}
In the region of frequency we are primarily interested in throughout this work,
\(
\omega \simeq E_{g},
\)
the sum in~\eqref{eqn:ImSigmadef} is dominated by states lying close to the band-edge critical point $\bm{k}^*$, as is shown explicitly for the checkerboard insulator in~\cref{app:cDOSLog}.  
Assuming that within this region the coupling changes only weakly, we are allowed to neglect its explicit $\bm{k}$-dependence~\cite{BassaniPastoriParravicini1975} and set \(
g \equiv g(\bm{k}^*)\neq 0\).
Then, the contribution to the imaginary part of the retarded self-energy is determined by the JDOS defined in~\cref{eqn:JDOS_def},
as made explicit in~\cref{eq:ImSigma}.
Once $\Sigma^R_{\mathrm{ph}}(\omega)$ is known, the polaritonic dispersion and spectral width are obtained by solving the secular equation $\det\{\bigl[\mathcal{D}_{\mathrm{ph}}^{R}(\omega)\bigr]^{-1}\} = 0$.
However, since the effective action has been truncated at quadratic order in the photon fields, neglecting quartic corrections arising from the effective four-photon interaction vertex [\cref{fig:neglected_corrections}(b)], the resulting dispersions are perturbatively accurate only up to second order in the light–matter coupling $g$.
One can then also safely neglect the contribution coming from off-diagonal terms in the Nambu structure of \cref{eqn:Nambu_Dinv},  represented in \cref{fig:neglected_corrections}(a), because it contributes only to fourth order in the coupling. Performing this approximation is equivalent to neglect in the effective action all terms that violate photon-number conservation and to work {only} with {the} retarded photonic propagator $D^{R}_{\mathrm{ph}}(t)= -\,\mathrm i\,\theta(t)\,\bigl\langle[\hat a(t),\hat a^{\dagger}(0)]\bigr\rangle,$ as was done in~\cref{sec:Van_Hove_Pol}. 

\section{Analytical derivation of JDOS for the checkerboard lattice}
\label{app:cDOSLog}
This appendix presents two complementary analytical routes to derive the singularity in the JDOS of the checkerboard lattice at the gap [see \cref{eqn:Div_Rho}]. 
First, we work directly with the exact tight-binding dispersion, arriving at a closed-form expression that is valid across the entire BZ.
Second, we focus on the low-energy physics near the M point and employ the Taylor expansion of the bands given in~\cref{eqn:Energy_Bands}: this highlights the connection between the 
M point and the singular behavior displayed in \cref{eqn:Div_Rho}.
\subsection{Derivation from the full Tight-Binding Dispersion}
\begin{figure}[tb]
    \centering
\includegraphics[width=\linewidth]{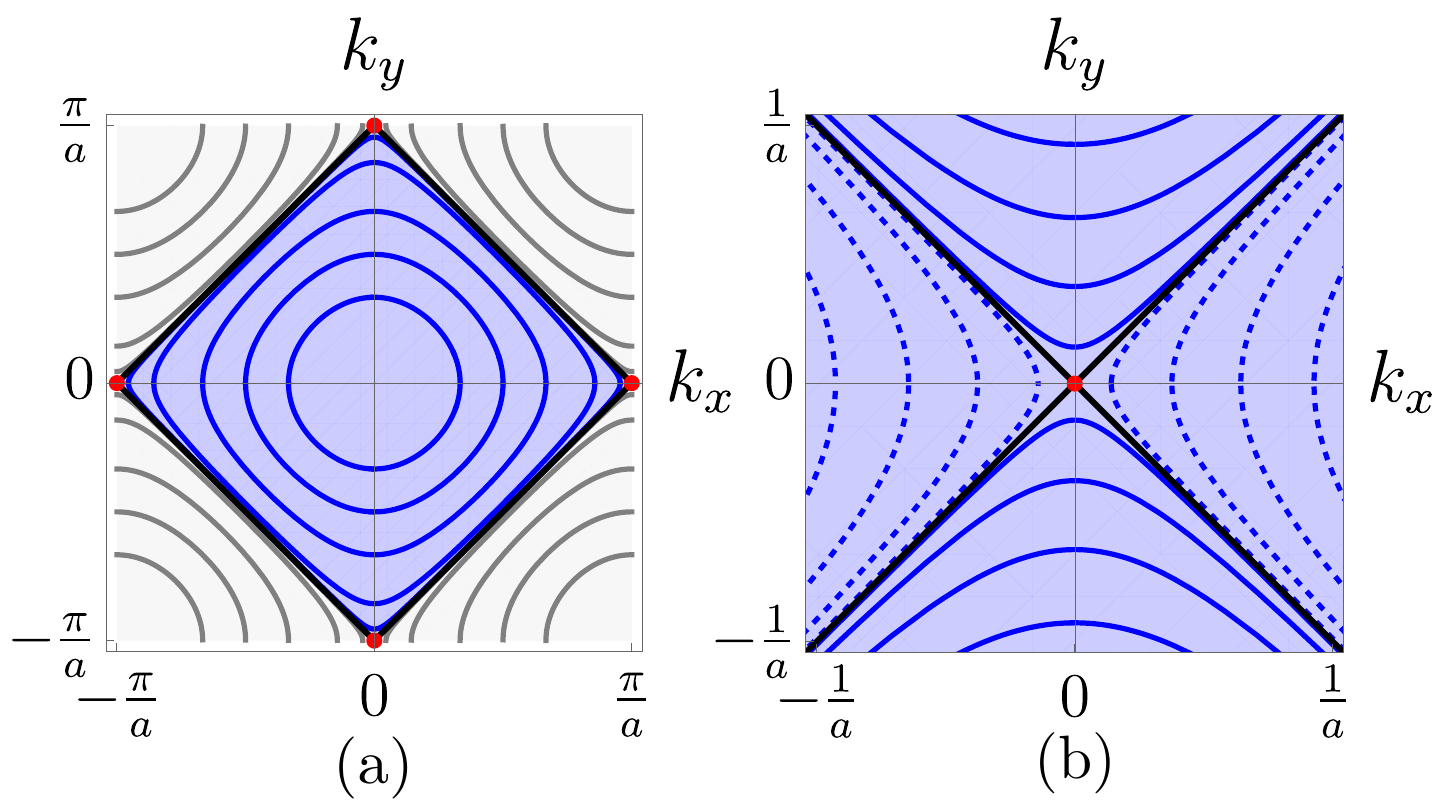}
\caption{(a) Constant-energy contours of the NN tight-binding dispersion on a square lattice,
$\displaystyle \varepsilon(k_x,k_y)=-2t\,[\cos(k_x a)+\cos(k_y a)]$ inside the corresponding first BZ $\mathcal{Q}=\{\abs{k_x},\abs{k_y}<\pi/a\}$. The blue shaded zone is the reduced BZ zone for the checkerboard and red dots indicate the $M$ point and its images at the four vertices.
(b) Constant-energy contours of the Taylor-expanded band-gap dispersion of the checkerboard lattice, $\delta E(\bm{k}) = E_+(\bm{k}) - E_-(\bm{k})$, plotted with momenta measured from $\bm{k}^{*} = (\pi/a,0)$. Blue lines correspond to energies $\delta E_{\bm{k}} > E_g$; the black line corresponds to the $\bm{k}$-points exactly at the gap, $\delta E_{\bm{k}} = E_g$.}
\label{fig:BZ_together}
\end{figure}
For the tight-binding checkerboard lattice introduced in~\cref{sec:cVHS}, the exact band energies are given in~\cref{eqn:bands_tight}. Inserting those dispersions into the definition~\eqref{eqn:JDOS_def} of the JDOS and taking the continuum limit of the quasi-momentum sum yields:
\begin{equation}
J(\omega)=\int_{\mathcal D}\frac{d\bm{k}}{(2\pi)^2}\,
\delta\left(\omega-2\sqrt{\varepsilon_{\bm{k}}^2+(E_g/2)^2}\right).
\label{eqn:JDOS_def_app}
\end{equation}
Here, $\mathcal{D}=\{\lvert k_x\rvert+\lvert k_y\rvert\le\pi/a\}$ denotes the diamond-shaped reduced BZ, illustrated in~\cref{fig:BZ_together}(a), inside the full BZ of the square-lattice tight-binding dispersion $\varepsilon_{\bm{k}}=-2t\,[\cos(k_x a)+\cos(k_y a)]$ in the absence of a staggered potential. For a real function $F(s)$ with isolated simple zeros $s_i$, one has the standard identity
\begin{equation}
\delta\bigl(F(\varepsilon)\bigr)=\sum_i\frac{\delta(\varepsilon-\varepsilon_i)}{\lvert F'(\varepsilon_i)\rvert}.
\label{eqn:delta_id}
\end{equation}
Choosing $F(\varepsilon)=\omega-2\sqrt{\varepsilon^2+(E_g/2)^2}$, the solutions of $F(\varepsilon)=0$ are
\begin{equation}
\varepsilon_\pm=\pm{\sqrt{\omega^{2}-E_g^{2}}}\Big{/}{2}
\qquad \text{for } \omega\ge E_g ,
\label{eqn:roots}
\end{equation}
and the derivatives evaluate to $F'(\varepsilon_{\pm})
           =-\,\frac{4\varepsilon_{\pm}}{\omega}$
 .
Applying the identity \eqref{eqn:delta_id} to \cref{eqn:JDOS_def_app} yields
\begin{equation}
J(\omega)=
\frac{\theta(\omega-E_g)\,\omega}{2\sqrt{\omega^{2}-E_g^{2}}}
\!\int_{\mathcal D}\!\frac{d\bm{k}}{(2\pi)^2}\,\delta\Bigl(\!\varepsilon_{\bm{k}}+{\sqrt{\omega^{2}-E_g^{2}}}\Big{/}{2}\Bigr).
\label{eqn:JDOS_single_root}
\end{equation}
Note that only the contribution from the negative root,
$\varepsilon_- = -\sqrt{\omega^{2}-E_g^{2}}/2$, survives. This happens because the negative branch energy contours $\varepsilon = s_-$ [blue curves in~\cref{fig:BZ_together}(a)] lie entirely inside the diamond $\mathcal D$, whereas the positive branch energy contours $\varepsilon = s_+$ [grey curves in~\cref{fig:BZ_together}(a)] fall outside this region and therefore do not contribute to the integral.
\cref{eqn:JDOS_single_root} can then be recast in terms of the square-lattice density of states,
\begin{equation}
\rho_\varepsilon(E)=\frac{1}{(2\pi)^2}\int_{\mathcal Q}d\bm{k}\,
\delta\bigl(E-\varepsilon_{\bm{k}}\bigr),
\label{eqn:DOS_Metal}
\end{equation}
where $\mathcal{Q}=\{\abs{k_x},\abs{k_y}<\pi/a\}$  denotes the full BZ, yielding
\begin{equation}
J(\omega)=
\frac{\theta(\omega-E_g) \,\omega }{2\sqrt{\omega^{2}-E_g^{2}}}
\rho_\varepsilon\Bigl(-{\sqrt{\omega^{2}-E_g^{2}}}\Big{/}{2}\Bigr).
\label{eqn:factorizedl}
\end{equation}
Although \cref{eqn:DOS_Metal} formally integrates over the entire square BZ, the negative argument
$-\sqrt{\omega^{2}-E_g^{2}}/2$ restricts the support to states lying within the diamond $\mathcal D$, thereby making~\cref{eqn:factorizedl} exactly equivalent to~\cref{eqn:JDOS_single_root}.

The density of states for a NN tight-binding band on a square lattice is standard textbook material (see Ref.~\cite{EconomouBook} for a derivation) and reads
\begin{equation}
  \rho_\varepsilon(E)=\frac{1}{2\pi^{2} a^2t }\,
                      K\!\left(\sqrt{1-(E/4t)^{2}}\right)\,
                      \theta(4t-|E|),
  \label{eqn:DOS_Elliptic}
\end{equation}
where $K$ denotes the complete elliptic integral of the first kind.  Inserting \cref{eqn:DOS_Elliptic} into the factorized expression~\cref{eqn:factorizedl} gives the closed form
\begin{equation}
  J(\omega)=
            \frac{\theta(\omega-E_g)\,\omega}{4\pi^{2}a^2t\sqrt{\omega^{2}-E_g^{2}}}\,
            K\!\left(\sqrt{1-\frac{\omega^{2}-E_g^{2}}{64t^{2}}}\right).
  \label{eqn:Closed_Form}
\end{equation}
For small energies $\lvert E\rvert\ll4t$ one has the expansion
\begin{equation}
  K\!\left(\sqrt{1-(E/4t)^{2}}\right)
  =-\log \!\bigl(\lvert E\rvert/4t\bigr)+\mathcal O\!\bigl((E/4t)^{2}\bigr),
  \label{eqn:K_smallE}
\end{equation}
so that, setting $E=-\sqrt{\omega^{2}-E_g^{2}}\Big{/}{2}$ as dictated by~\cref{eqn:factorizedl}, one immediately recovers the additional logarithmic divergence of \cref{eqn:Div_Rho}.  The logarithmic enhancement is therefore traced back to the conventional 2D VHS associated with the saddle point of the square lattice dispersion $\varepsilon_{\bm{k}}$ at the point $\bm{k}^*=(\pi/a,0)$ \cite{VanHove1953}.

\subsection{Derivation from the Taylor-expanded dispersion around the M point}
Inserting the energy bands expanded around the critical point \(\bm{k}^*\) given in~\cref{eqn:Energy_Bands} into the definition of the JDOS, \cref{eqn:JDOS_def}, yields
\begin{equation}\label{eqn:JDOS_def_cont}
J(\omega)= \!\!\int_{|\bm{k}|<\Lambda} \!\!\!\!\frac{dk_x\,dk_y}{(2\pi)^2}\,
          \delta\!\Bigl(\omega-E_g-2\gamma\bigl[(k_xa)^2-(k_ya)^2\bigr]^2\Bigr),
\end{equation}
where, through the shift \(\bm{k}\!\to\!\bm{k}-\bm{k}^*\), we have centered the reference frame at $\bm{k}^*$ and introduced a momentum cutoff \(\Lambda\) to model the finite bandwidth. 
It is convenient to recast the JDOS in the form
\begin{equation}\label{eqn:adi_jdos}
J(\omega)=\frac{1}{2a^{2}\gamma}\,\rho(\tilde{\epsilon})\Big{|}_{\tilde{\epsilon}=(\omega-E_g)/(2\gamma)},
\end{equation}
 where the dimensionless density of states is defined as
\begin{equation}
\rho(\tilde{\epsilon})
       =\int_{|\bm{x}|<\tilde{\Lambda}}
        \frac{dx\,dy}{(2\pi)^2}\,
        \delta\bigl(\tilde{\epsilon}-\tilde{\epsilon}(x,y)\bigr),
\end{equation}
and involves the dimensionless dispersion
\begin{equation}
\tilde{\epsilon}(x,y)=(x^{2}-y^{2})^{2}.
\end{equation}
Here we introduced the dimensionless variables $x = k_x a$ and $y = k_y a$, together with the dimensionless cutoff $\tilde{\Lambda} = \Lambda a$ and the dimensionless energy $\tilde{\epsilon}=\epsilon/(2 \gamma)$.
The Dirac delta confines the integration to the constant-energy contour 
\begin{equation}
S(\tilde\epsilon)
  =\bigl\{(x,y):\tilde\epsilon(x,y)=\tilde\epsilon,\;|\bm{x}|\le\tilde\Lambda\bigr\},  
\end{equation}
giving the line integral
\begin{equation}
\rho(\tilde\epsilon)
  =\frac{1}{4\pi^{2}}
   \oint_{S(\tilde\epsilon)}
   \frac{dl}{\bigl|\nabla_{\!\bm{x}}\tilde\epsilon\bigr|}.    
\end{equation}
Here $l$ is the arclength parameter along $S(\tilde\epsilon)$.
For $\tilde\epsilon(x,y)=(x^{2}-y^{2})^{2}$ the condition $\tilde\epsilon(x,y)=\tilde\epsilon$ reduces to
$
x^{2}-y^{2}=\pm\tilde\epsilon^{1/2}$. Hence the full contour decomposes into two branches:
\begin{equation}
\begin{aligned}
S_{1}(\tilde\epsilon)
   &=\Bigl\{(x,y):\;
        y=\pm\sqrt{x^{2}+\tilde\epsilon^{1/2}},
        \;|\bm{x}|\le\tilde\Lambda\Bigr\},\\[6pt]
S_{2}(\tilde\epsilon)
   &=\Bigl\{(x,y):\;
        y=\pm\sqrt{x^{2}-\tilde\epsilon^{1/2}},\;
        |x|\ge\tilde\epsilon^{1/4},\;
        |\bm{x}|\le\tilde\Lambda\Bigr\}.
\end{aligned}
\end{equation}
Thus $S(\tilde{\epsilon})=S_{1}(\tilde{\epsilon})\cup S_{2}(\tilde{\epsilon})$.
Consequently, the density of states separates into two contributions,
\[
\rho(\tilde{\epsilon})=\rho_{1}(\tilde{\epsilon})+\rho_{2}(\tilde{\epsilon}),
\qquad
\rho_{i}(\tilde{\epsilon})=\frac{1}{4\pi^{2}}
   \!\!\oint_{S_i(\tilde\epsilon)}
   \!\!\frac{dl}{\bigl|\nabla_{\!\bm{x}}\tilde\epsilon\bigr|}\;(i=1,2).
\]
The subset $S_{1}$ has two branches with vertices at
$(0,\pm\tilde{\epsilon}^{1/4})$ that open vertically
(upward and downward), whereas $S_{2}$ has two branches with vertices at
$(\pm\tilde{\epsilon}^{1/4},0)$ that open horizontally
(leftward and rightward).
\cref{fig:BZ_together}(b) depicts these contours as solid and dashed blue lines, respectively. Note that all branches correspond to $\tilde{\epsilon}>0$.
  In the original variables, they correspond to $\omega>E_g$.
 As $\tilde{\epsilon}$ approaches zero, the blue contours shrink until, at $\tilde{\epsilon}=0$ ($\omega=E_g$), they become two diagonally crossing critical lines (black lines in~\cref{fig:BZ_together}(b)) intersecting at $(0,0)$, i.e.~the point $\bm{k}=\bm{k}^*$ in the shifted reference frame. Notice that for $\tilde{\epsilon}<0$ no constant-energy contours exist, reflecting the absence of available states below the gap.
 
Since the dispersion \(\tilde{\epsilon}(x,y)=(x^{2}-y^{2})^{2}\) and the integrand
\(dl/|\nabla\tilde{\epsilon}|\) are invariant under the exchange
\(x\leftrightarrow y\), which maps \(S_{1}(\tilde{\epsilon})\) onto
\(S_{2}(\tilde{\epsilon})\), the two contour integrals are identical,
\[\,\rho_{1}(\tilde{\epsilon})=\rho_{2}(\tilde{\epsilon}),
\]
so it suffices to evaluate only one of the two. We choose $\rho_{1}(\tilde{\epsilon})$.
Along the upper branch of $S_{1}(\tilde{\epsilon})$ we may write
$y(x)=\sqrt{x^{2}+\sqrt{\tilde{\epsilon}}}$ with $|x|\le\tilde{\Lambda}$. The arc-length element is
$dl=\sqrt{1+(dy/dx)^{2}}\,dx       =\sqrt{(2x^{2}+\sqrt{\tilde{\epsilon}})/(x^{2}+\sqrt{\tilde{\epsilon}})}\,dx$,
whereas on the same contour
$|\nabla\tilde{\epsilon}|=4\sqrt{\tilde{\epsilon}}\sqrt{2x^{2}+\sqrt{\tilde{\epsilon}}}$.
Their ratio therefore becomes
$$
\frac{dl}{|\nabla\tilde{\epsilon}|}
     =\frac{dx}{4\sqrt{\tilde{\epsilon}}\sqrt{x^{2}+\sqrt{\tilde{\epsilon}}}}\;.
$$
Because the integrand is even in $x$ and the lower branch ($y\!\to\!-y$) contributes identically, the integral over the whole subset $S_{1}$ gains an overall factor $4$.  Substituting into the definition, one obtains
$$
2\rho_{1}(\tilde{\epsilon})
   =\frac{2}{4\pi^{2}}
     \int_{S_{1}(\tilde{\epsilon})}\frac{dl}{|\nabla\tilde{\epsilon}|}
   =\frac{2}{4\pi^{2}\sqrt{\tilde{\epsilon}}}
     \int_{0}^{\tilde{\Lambda}}
     \frac{dx}{\sqrt{x^{2}+\sqrt{\tilde{\epsilon}}}},
$$
where we note the inverse square‐root divergence as \mbox{$\tilde{\epsilon}\to 0$}. The remaining integral is elementary, giving
\[
\rho(\tilde{\epsilon})=2\rho_{1}(\tilde{\epsilon})
   =\frac{2}{4\pi^{2}\sqrt{\tilde{\epsilon}}}\,
     \operatorname{arcsinh}\!\bigl(\tilde{\Lambda}\,\tilde{\epsilon}^{-1/4}\bigr),
\] For $\tilde{\epsilon}\to0$ this behaves as
$\rho(\tilde{\epsilon})\simeq\frac{1}{2\pi^{2}\sqrt{\tilde{\epsilon}}}
      \bigl(-\tfrac14\ln|\tilde{\epsilon}|\bigr)$,
thereby making the logarithmic enhancement explicit. Finally restoring the dimensionful JDOS through~\cref{eqn:adi_jdos} reproduces the behavior shown in~\cref{eqn:Div_Rho}.
\section{Analytical derivation of \(\Re[\Sigma^R_{\mathrm{ph}}(\omega)]\) for \(\omega \sim E_g^{-}\) }\label{app:deriv.Re}
In \cref{fig:Self} we plotted the curves $\Im\Sigma_{\mathrm{ph}}^R(\omega)$ and $\Re\Sigma_{\mathrm{ph}}^R(\omega)$ corresponding to three insulating cases: the 1D and 2D parabolic bands, and the checkerboard lattice.  We have commented the singular behaviors of both quantities at the band-gap energy $E_g$ in all three cases. These behaviors are summarized in the first two columns of \cref{tab:sigma_behaviors}. We recall that the features observed in $\Im\Sigma_{\mathrm{ph}}^R(\omega)$ are the same of the JDOS $J(\omega)$ due to the relation in~\cref{eq:ImSigma}. The forms followed by $\Im\Sigma_{\mathrm{ph}}^R(\omega)$ in the two parabolic cases are standard textbook results~\cite{grosso2013solid}. The behavior of $\Im\Sigma_{\mathrm{ph}}^R(\omega)$ (and of $J(\omega)$) in the checkerboard case, shown in~\cref{eqn:Div_Rho}, was already justified in~\cref{app:cDOSLog}.
In this appendix, we derive the behavior of $\Re\Sigma_{\mathrm{ph}}^R(\omega)$ in the limit $\omega\rightarrow E_g^{-}$ for the three cases of \cref{tab:sigma_behaviors}. 
These expressions can be obtained by applying the Kramers--Kronig relation given in~\cref{eq:ReSigma}, assuming that the function $J(\omega)$ is known for each case. For a subgap frequency $\omega= (1- \epsilon) E_g$ with $\epsilon > 0$, the pole lies strictly outside the integration domain, so no Cauchy principal value is needed, and we have the expression
\begin{equation}\label{eqn:Math_Anal}
    \Re\Sigma^R_{\mathrm{ph}}(\epsilon) \propto I(\Lambda,\varepsilon)\equiv \int_{1+\delta}^{\Lambda}\! dx
      \frac{x\,\tilde{J}(x)}
           {x^{2}-\bigl(1-\epsilon\bigr)^{2}},
\end{equation}
 where all the frequency dependence of $J(\omega)$ has been encoded in the dimensionless function $\tilde{J}(x)$ and we introduced the dimensionless variable $x = \omega/E_g$ and a cut-off $\Lambda$ to model the physical bandwidth limit. The behavior of the integral $I(\Lambda,\varepsilon)$ in~\eqref{eqn:Math_Anal} in the limit $\varepsilon\rightarrow0^+$  can be obtained analytically for each of the three cases.

We first carry out the calculation explicitly for the checkerboard case, for which $\tilde{J}(x)=\log(x-1)/\sqrt{x-1}$.
Making the substitution $x=1+t^{2}\,\bigl(dx=2t\,dt\bigr)$ with $t\in[0,\sqrt{\Lambda-1}]$ removes the square root and casts the integral in the form  \begin{equation}\label{eqn:split_integral_cheq} 
I_{\text{ch}}(\Lambda,\varepsilon)=-2\int_{0}^{\sqrt{\Lambda-1}}\!\!\!\!dt \Bigl(\frac{\log t}{t^{2}+\varepsilon}+\frac{\log t}{t^{2}+2-\varepsilon}\Bigr). 
\end{equation}  
The only term that can diverge as $\varepsilon\to0^{+}$ is the first one. Because the singularity comes exclusively from the \mbox{small-$t$} region, we can extend its upper limit to $\infty$ without changing the leading behavior:  
\begin{equation} 
I_{\text{ch}}(\Lambda,\varepsilon)\;\mathrel{\overset{\varepsilon \to 0^{+}}{\simeq}}\;       -2\!\int_{0}^{\infty}\!\!\! dt \,\frac{\log t}{t^{2}+\varepsilon}      = \frac{\pi}{2}\,\frac{-\log\varepsilon}{\sqrt{\varepsilon}}, 
\end{equation} 
where, to obtain the final equality, we introduced the new integration variable $u=t/\sqrt{\varepsilon}$, recongnized the indefinite integral $\int \!\tfrac{du}{1+u^{2}}=\arctan{u}$ and employed the identity $\int_{0}^{\infty}\!\tfrac{\log u}{1+u^{2}}\,du=0$. To recover the full value of $I_{\text{ch}}(\Lambda,\varepsilon)$ one should subtract the finite “tail’’ $\int_{\sqrt{\Lambda-1}}^{\infty}\!\log t/(t^{2}+\varepsilon)\,dt$ that was introduced by extending the upper limit to $\infty$, and add the second, always-convergent integral appearing in~\cref{eqn:split_integral_cheq}.  Retaining these pieces yields an additional, $\Lambda$-dependent constant that follows the leading term and remains finite as $\varepsilon\to0^{+}$.  Hence 
\begin{equation} I_{\text{ch}}(\Lambda,\varepsilon)=\frac{\pi}{2}\,\frac{-\log\varepsilon}{\sqrt{\varepsilon}}+\mathcal O(1). 
\end{equation}
For the remaining two cases the calculation proceeds analogously to the checkerboard example.
In the parabolic 1D case, where $\tilde J(x)=1/\sqrt{x-1}$, rewriting the integral in terms of the variable $t$ splits it into two parts:
\begin{equation}
I_{1\text{D}}(\Lambda,\varepsilon)=\int_{0}^{\sqrt{\Lambda-1}}\!dt
        \Bigl(\frac{1}{t^{2}+\varepsilon}+ \frac{t^{2}}{t^{2}+\varepsilon}\Bigr).
\end{equation}
Extending the integration of the first term to $\infty$ one finds
\begin{equation}
I_{1\text{D}}(\Lambda,\varepsilon)=\frac{\pi}{2\sqrt{\varepsilon}}+\mathcal O(1),
\label{eqn:I_1D}
\end{equation}
having once again substituted $t=u\sqrt{\varepsilon}$ and recognized the same indefinite integral used in the checkerboard case. We note that a result similar to \cref{eqn:I_1D} has been already obtained in \cite{Islam2022}. Finally, in the parabolic 2D case, for which $\tilde J(x)=1$, we have
\begin{equation}
I_{2\text{D}}(\Lambda,\varepsilon)=\int_{0}^{\sqrt{\Lambda-1}}\! \!\!\!dt \Bigl(\frac{t}{t^{2}+\varepsilon}+\frac{t^{3}}{t^{2}+\varepsilon}\Bigr).
\end{equation}
Proceeding as before, after the integration we obtain
\begin{equation}
I_{2\text{D}}(\Lambda,\varepsilon)
=
-\frac12\ln\bigl(\varepsilon \bigr)
+\mathcal O(1),
\end{equation}
having recognized the indefinite integral $\int \!\frac{t\,dt}{t^{2}+\varepsilon}= \tfrac12\ln\!\bigl(t^{2}+\varepsilon\bigr)$.
Upon returning to the original frequency variable $\omega$, one recovers the singular behaviors reported in the second column of~\cref{tab:sigma_behaviors}.
\section{Analytical derivation of the polariton shift scaling laws}\label{app:deriv.branche}
In this section, we present an analytical derivation of the scaling behavior of the polariton shift $\delta\omega_\text{P} = \omega_0 - \omega_\text{P}$ as a function of the dimensionless light-matter coupling strength $\tilde{g} = g/E_g$ in the weak-coupling regime. These scaling laws $\delta\omega_p(\tilde{g})$ are presented in \cref{fig:Branch} and summarized in the third column of~\cref{tab:sigma_behaviors}. We obtained these results by solving the equation
\[
\Re[D^R_{\mathrm{ph}}(\omega)]^{-1} =
\omega - \omega_0 - \operatorname{Re}\bigl[\Sigma^{R}_{\mathrm{ph}}(\omega)\bigr] = 0,
\]
with $\omega_0 = E_g$, in the limit $\omega \rightarrow E_g^-$.
In this regime, it is justified to approximate $\operatorname{Re}\bigl[\Sigma^{R}_{\mathrm{ph}}(\omega)\bigr]$ around $\omega \sim E_g^+$ (see the second column of~\cref{tab:sigma_behaviors}), leading to the following equations: 
\begin{align}
\delta\tilde{\omega}_\text{P} &= \tilde{g}^2 a_{\text{1D}} \frac{1}{\sqrt{\delta\tilde{\omega}_\text{P}}}, \\
\delta\tilde{\omega}_\text{P} &= -\tilde{g}^2 a_{\text{2D}}\log\delta\tilde{\omega}_\text{P}, \\
\delta\tilde{\omega}_\text{P} &= -\tilde{g}^2 a_{\mathrm{ch}} \frac{\log\delta\tilde{\omega}_\text{P}}{\sqrt{\delta\tilde{\omega}_\text{P}}},
\end{align}
where we introduced the dimensionless polariton shift $\delta\tilde{\omega}_{\text{P}}\equiv\delta\tilde{\omega}/E_{g}$ and, for each insulating phase, an overall constant prefactor $a$ that is independent of the light–matter coupling strength $g$. The equation for the 1D case can be solved analytically by inspection. For the remaining two cases, the solutions can be written in terms of the Lambert \( W \) function, defined as the solution of \( W(z)\,e^{W(z)} = z \).
The resulting expressions are: 
\begin{align}
\delta\tilde{\omega}_{P,\text{1D}} &=\tilde{g}^{\frac43} a^{\frac23}_{\text{1D}}, \label{eqn:delta_omega_1D} \\
\delta\tilde{\omega}_{P,\text{2D}} &= \tilde{g}^2 W\bigl((a_{\text{2D}}\tilde{g}^2)^{-1}\bigr)a_{\text{2D}} , \label{eqn:delta_omega_2D} \\
\delta\tilde{\omega}_{P,\text{ch}} &= \tilde{g}^{\frac43} \big[W\bigl(({2 a_{\text{ch}}\tilde{g}^2}/{3})^{-1}\bigr)\big]^{\frac23}\big({2 a_{\text{ch}}}/{3})^{\frac23} , \label{eqn:delta_omega_ch}
\end{align}
For weak coupling ($\tilde g\!\to\!0$), the Lambert function satisfies $W(z)\simeq \ln z - \ln\ln z + \dots$ for $z\gg 1$; using this expansion yields the leading-order scaling behaviors reported in the third column of~\cref{tab:sigma_behaviors}.

\section{Finite temperature corrections}\label{app:finiteT}

Starting from the general expression \cref{eqn:Self-Energy}, the imaginary part of retarded self energy follows from
$\lim_{\eta\to0^+}(x+i\eta)^{-1}=\mathrm{P.V.}(1/x)-i\pi\delta(x)$ and yields, at arbitrary temperature,
\begin{align}
\Im\Sigma^{R}_{\mathrm{ph}}(\omega;T)
= &-\!\frac{\pi}{N} \sum_{\bm{k}} g^2(\bm{k})\!
\bigl[n_F\!\bigl(E_-(\bm{k})\bigr)\!-\!n_F\!\bigl(E_+\!(\bm{k})\bigr)\!\bigr]\notag\\
&\quad\times\,
\Big[\delta\!\bigl(\omega-\delta E_{\bm{k}}\bigr)-\delta\!\bigl(\omega+\delta E_{\bm{k}}\bigr)\Big].
\label{eq:ImSigma_finiteT_k}
\end{align}
For the particle-hole-symmetric case considered in the main text,
$E_-(\bm{k})=-E_+(\bm{k})$ and $\mu=0$, one has
$
n_F\!\bigl(E_-(\bm{k})\bigr)-n_F\!\bigl(E_+(\bm{k})\bigr)
= \tanh\!\Big({\beta\,\delta E_{\bm{k}}}/{4}\Big).$
As in \cref{app:Deriv.Self}, close to the band-edge point $\bm{k}^*$ we can neglect the explicit
$\bm{k}$-dependence of the coupling and set $g(\bm{k})\simeq g(\bm{k}^*)\equiv g$.
Using the JDOS definition of \cref{eqn:JDOS_def}, the result can be written compactly as
\begin{align}
\Im\Sigma^{R}_{\mathrm{ph}}(\omega;T)
&= -\pi g^2 A_{\rm cell} \,
\tanh\!\Big(\frac{\beta\omega}{4}\Big)\,
\big[\,J(\omega)-J(-\omega)\,\big],
\label{eq:ImSigma_finiteT_JDOS}
\end{align}
which reduces to the $T=0$ expression \cref{eq:ImSigma} since $\tanh(\beta\omega/4)\to \mathrm{sgn}(\omega)$
for $\beta\to\infty$.
\cref{eq:ImSigma_finiteT_JDOS} shows that temperature enters $\Im\Sigma^{R}_{\mathrm{ph}}$
only through the smooth factor $\tanh(\beta\omega/4)$ that accounts for the population imbalance between valence and conduction
states. Consequently, in the vicinity of the gap one may approximate
$
\tanh(\frac{\beta\omega}{4})
\simeq \tanh (\frac{\beta E_g}{4}),
$
so the same scaling laws found at $T=0$ for $\Im\Sigma^{R}_{\mathrm{ph}}(\omega)$ at $\omega\to E_g^+$
remain valid at $T>0$, up to a temperature-dependent prefactor
$
\mathcal{F}_T \equiv \tanh(\frac{\beta E_g}{4}).
$
Using \cref{eq:ImSigma_finiteT_JDOS} one finds the finite-$T$ analogue of \cref{eq:ReSigma},
\begin{align}
&\nonumber \Re\Sigma^{R}_{\mathrm{ph}}(\omega;T) \\
&= -2g^2 A_{\rm cell}\,\mathrm{P.V.}\int_{0}^{\infty}\!d\omega'\,
\frac{\omega' J(\omega')}{\omega'^2-\omega^2}\,
\tanh\!\Big(\frac{\beta\omega'}{4}\Big).
\label{eq:ReSigma_finiteT}
\end{align}
For $\omega\to E_g^-$, the integral is dominated by frequencies $\omega'\gtrsim E_g$ where $J(\omega')$ is singular
(see \cref{app:deriv.Re}). Since $\tanh(\beta\omega'/4)$ is smooth at $\omega'=E_g$, to leading order one can pull it
out of the singular part of the integral by fixing again its value to $\mathcal{F}_T$:
\begin{equation}
\Re\Sigma^{R}_{\mathrm{ph}}(\omega;T)
\simeq \mathcal{F}_T\, \Re\Sigma^{R}_{\mathrm{ph}}(\omega;0)
\qquad (\omega\to E_g^-).
\label{eq:ReSigma_prefactor}
\end{equation}
As a consequence, finite temperature preserves the weak-coupling scaling of the polariton shift with respect to the dimensionless light--matter coupling $\tilde g=g/E_g$: temperature merely renormalizes the coupling-independent coefficients according to
\begin{equation}
a_{\text{1D}}\to \mathcal{F}_T\,a_{\text{1D}},\quad
a_{\text{2D}}\to \mathcal{F}_T\,a_{\text{2D}},\quad
a_{\text{ch}}\to \mathcal{F}_T\,a_{\text{ch}},
\label{eq:aT_replace}
\end{equation}
while leaving unchanged the weak-coupling exponents reported in \cref{tab:sigma_behaviors}. Explicitly, the solutions corresponding to \cref{eqn:delta_omega_1D,eqn:delta_omega_2D,eqn:delta_omega_ch} become
\begin{align}
\delta\tilde{\omega}_{P,\text{1D}}(T)
&= \tilde{g}^{\frac43}\bigl[\mathcal{F}_T a_{\text{1D}}\bigr]^{\frac23},
\label{eq:D4_T}\\
\delta\tilde{\omega}_{P,\text{2D}}(T)
&= \tilde{g}^2\,(\mathcal{F}_T a_{\text{2D}})\,
W\!\Big(\bigl(\mathcal{F}_T a_{\text{2D}}\tilde{g}^2\bigr)^{-1}\Big),
\label{eq:D5_T}\\
\delta\tilde{\omega}_{P,\text{ch}}(T)
&= \tilde{g}^{\frac43}\Big[W\!\Big(\bigl({2 \mathcal{F}_T a_{\text{ch}}\tilde{g}^2}/{3}\bigr)^{-1}\Big)\Big]^{\frac23}
\nonumber\\
&\quad\times \Big({2 \mathcal{F}_T a_{\text{ch}}}/{3}\Big)^{\frac23}.
\label{eq:D6_T}
\end{align}

\begin{figure}[tb]
\centering
\includegraphics[width=0.48\textwidth]{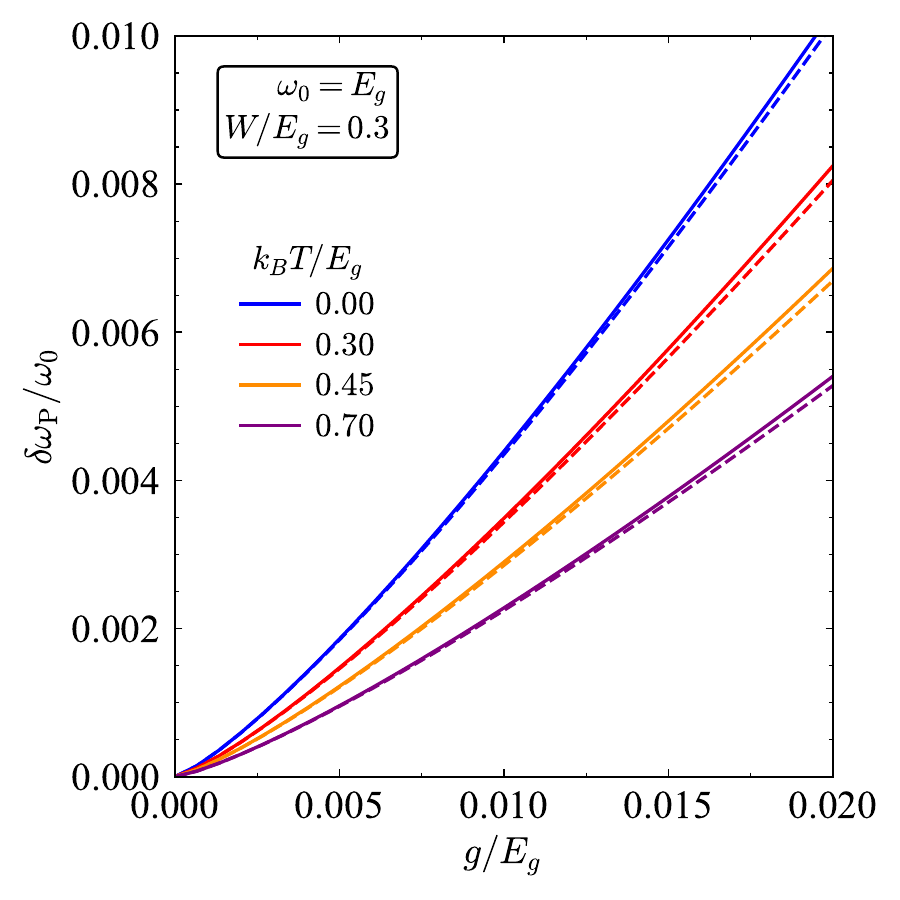}
\caption{Numerical (solid) and analytical (dashed) polariton shifts as a function of the dimensionless light--matter coupling $g/E_g$, for several values of the temperature $T$. Finite temperature suppresses the magnitude of the shift while preserving its weak-coupling scaling with respect to $g/E_g$.}
\label{fig:finiteT_shift_comparison}
\end{figure}

To illustrate these results, \cref{fig:finiteT_shift_comparison} compares the numerical polariton shift with the corresponding analytical expressions for several temperatures, including the zero-temperature limit. 
\Cref{fig:finiteT_shift_comparison} therefore provides a direct numerical confirmation of the finite-temperature analysis presented in this appendix: the analytical expressions in \cref{eq:D4_T,eq:D5_T,eq:D6_T} capture the thermal suppression of the polariton shift while leaving its asymptotic scaling unchanged.

\section{Subgap absorption}\label{app:subgapAbs}
Residual subgap absorption can generally originate from several mechanisms, including bound states such as excitons, disorder effects, finite-temperature corrections, or a finite fermionic lifetime. In the ultracold-atom implementation discussed in \cref{sec:Ultra-imp}, the first two mechanisms are absent. Excitonic bound states do not occur because the matter sector is realized with a gas of spin-polarized fermionic atoms, for which two-body interactions are absent. The single-particle band description is therefore exact. Disorder-induced subgap absorption is also absent in the implementation considered here: the atoms move in a clean optical lattice potential, with negligible static random impurities and lattice disorder.
Finite temperature does not open a subgap absorption channel either. As shown in \cref{app:finiteT}, the finite-temperature interband response is modified only through the thermal population imbalance between valence and conduction bands, while the absorption threshold remains fixed by the band gap.
The only possible generic source of residual subgap absorption is then a finite fermionic lifetime, i.e.~a broadening of the fermionic single-particle spectral functions. In the first part of this appendix, \cref{appsub:bubble_finite}, we show how a constant fermionic broadening produces residual subgap absorption and affects the lower-polariton energy shift relative to the zero-absorption case. In the second part, \cref{appsub:fock}, we estimate this broadening microscopically from the cavity-mediated one-loop Fock self-energy, finding it to be too small to appreciably affect the subgap polariton branch for realistic parameters.

\subsection{Fermionic spectral broadening corrections}\label{appsub:bubble_finite}
To assess the effect of a finite and constant fermionic spectral broadening, we recompute the photonic self-energy in Matsubara frequencies starting from \cref{eqn:Self-Energy}, but replacing the bare fermionic propagators with damped ones. For a fermion in band \(\lambda=\pm\), we introduce the retarded propagator
\begin{align}
G_{\lambda}^{R}(\bm{k},\varepsilon)
=
\frac{1}{\varepsilon-E_{\lambda}(\bm{k})+i\eta_{\lambda}},
\end{align}
where \(\eta_{\lambda}>0\) is the spectral broadening, or inverse lifetime, of a fermion in band \(\lambda\).
In Matsubara space, the corresponding propagator can be written in spectral form as
\begin{align}
{G}_{\lambda}(\bm{k},\omega_n)
=
\int_{-\infty}^{\infty}\frac{d\varepsilon}{2\pi}
\frac{\rho_{\lambda}(\bm{k},\varepsilon)}{i\omega_n-\varepsilon},
\end{align}
with the Lorentzian-shaped fermionic spectral density
\begin{align}
\rho_{\lambda}(\bm{k},\varepsilon)
=
-2\,\Im G_{\lambda}^{R}(\bm{k},\varepsilon)
=
\frac{2\eta_{\lambda}}{\bigl[\varepsilon-E_{\lambda}(\bm{k})\bigr]^2+\eta_{\lambda}^2}.
\label{eq:fermionic_spectral_eta}
\end{align}
At zero temperature, after analytic continuation \(i\omega_{n}\to \omega+i0^+\), the imaginary part of the retarded photon self-energy can be written as
\begin{align}
\Im\Sigma_{\rm ph}^{R}(\omega)
&=
-\frac{1}{2N}\sum_{\bm k} g(\bm k)^2
\int\frac{d\varepsilon}{2\pi}\,
W_\omega(\varepsilon)\notag\\&\times
\Big[
\rho_{-}(\bm k,\varepsilon)\rho_{+}(\bm k,\varepsilon+\omega)\\&\quad\,\,
+
\rho_{+}(\bm k,\varepsilon)\rho_{-}(\bm k,\varepsilon+\omega)
\Big],
\end{align}
where \(W_\omega(\varepsilon)\equiv \theta(-\varepsilon)-\theta(-\varepsilon-\omega)\) is a factor that keeps into account the zero-temperature fermionic distributions $n_F(\varepsilon)=\theta(-\varepsilon)$.
For \(\omega>0\), the term \(\rho_{-}(\bm k,\varepsilon)\rho_{+}(\bm k,\varepsilon+\omega)\) is resonant because the two spectral peaks overlap when
\(\varepsilon\simeq E_{-}(\bm k)\) and \(\varepsilon+\omega\simeq E_{+}(\bm k)\), namely for
\(\omega\simeq \delta E_{\bm k}\equiv E_{+}(\bm k)-E_{-}(\bm k)\).
In contrast, \(\rho_{+}(\bm k,\varepsilon)\rho_{-}(\bm k,\varepsilon+\omega)\) would require
\(\omega\simeq E_{-}(\bm k)-E_{+}(\bm k)=-\delta E_{\bm k}<0\) to be resonant, and therefore has no on-shell support for positive \(\omega\).
Its contribution is thus controlled only by the off-shell Lorentzian tails and is negligible when \(\eta_\lambda\ll E_g\).
Near resonance, \(\omega\simeq \delta E_{\bm k}\), the product \(\rho_{-}(\bm k,\varepsilon)\rho_{+}(\bm k,\varepsilon+\omega)\) is non-negligible only for \(\varepsilon\) within a narrow interval of width of the order of \(\eta_\Sigma\equiv \eta_-+\eta_+\) around \(\varepsilon\simeq E_-(\bm k)\). Since \(E_-(\bm k)\le -E_g/2\) and \(E_-(\bm k)+\omega\simeq E_+(\bm k)\ge E_g/2\), this interval lies a distance of at least \(E_g/2\) away from both edges of the domain \((-\omega,0)\) where \(W_\omega(\varepsilon)=1\). Therefore, for \(\eta_\Sigma\ll E_g\) and $\omega\simeq \delta E_{\bm{k}}$, one may replace \(W_\omega(\varepsilon)\) by \(1\) over the support of the overlap and obtain
\begin{align}
\Im\Sigma_{\rm ph}^{R}(\omega)
\simeq
-\frac{\eta_{\Sigma}}{N}\sum_{\bm{k}}g(\bm{k})^2\,
\frac{1}{\bigl(\omega-\delta E_{\bm{k}}\bigr)^2+\eta_{\Sigma}^{\,2}}.
\label{eq:ImSigma_subgap_final}
\end{align}
This shows that the sharp energy-conservation delta peak \(\delta(\omega-\delta E_{\bm k})\) of the zero-broadening result of \cref{eqn:ImSigmadef} is replaced by a Lorentzian of width \(\eta_\Sigma\), given by the sum of the fermionic broadenings $\eta_-$ and $\eta_+$. 
\begin{figure}[tb]
    \centering
\includegraphics[width=1.0\linewidth]{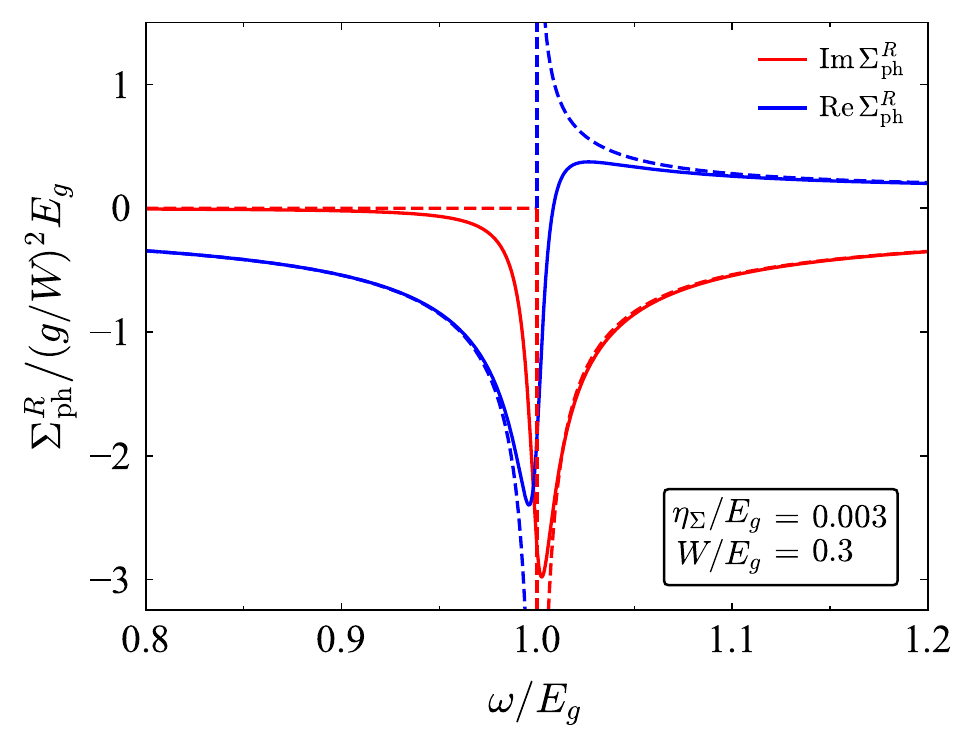}
\caption{\(\Im\Sigma_{\rm ph}^R(\omega)\) and
\(\Re\Sigma_{\rm ph}^R(\omega)\) computed for a representative value of the spectral broadening \(\eta_\Sigma/E_g=0.003\). The corresponding zero-damping results are shown as dashed curves for
comparison. A finite \(\eta_\Sigma\) rounds the threshold singularity at
\(\omega\simeq E_g\) and generates a weak subgap absorption tail.}
\label{fig:Self_finite_life}
\end{figure}
\begin{figure}[tbh]
    \centering
\includegraphics[width=0.9\linewidth]{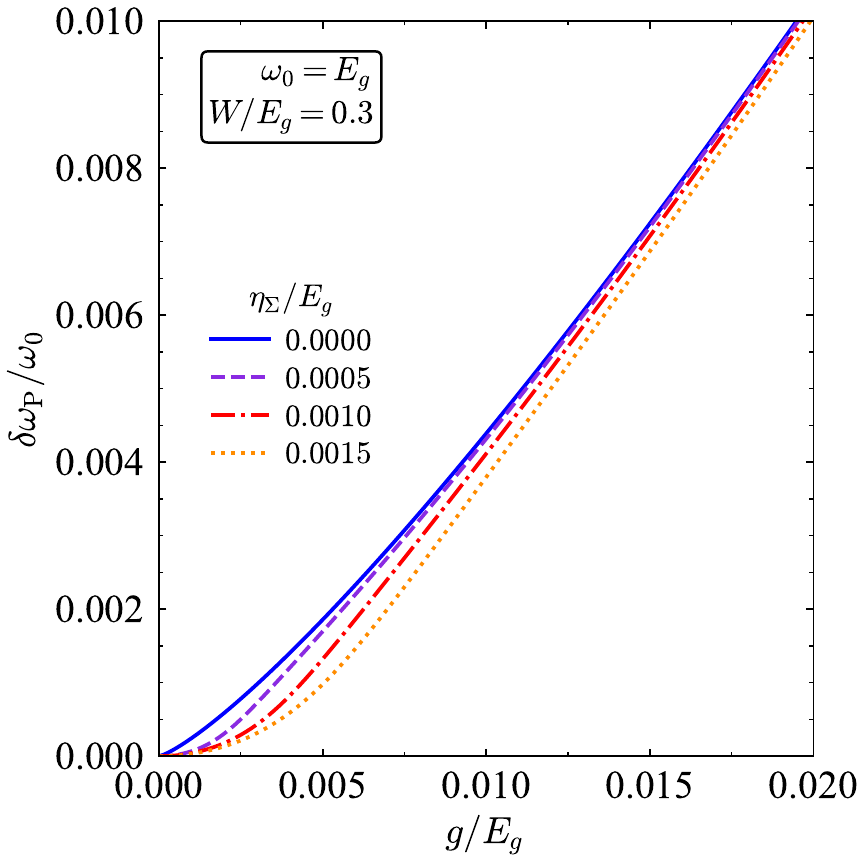}
\caption{
Lower polariton shift $\delta \omega_{p}$ as a function of \(g/E_g\) for increasing values of the
total fermionic broadening \(\eta_\Sigma/E_g\). 
At fixed \(\eta_\Sigma>0\) and upon increasing the coupling \(g\), the shift is first suppressed but then recovers the  \(\eta_\Sigma=0\) behavior.}
\label{fig:pol_shift_finite_life}
\end{figure}

In \cref{fig:Self_finite_life} we plot \(\Im\Sigma_{\rm ph}^R(\omega)\) and
\(\Re\Sigma_{\rm ph}^R(\omega)\) for a representative broadening
\(\eta_\Sigma/E_g=3\times 10^{-3}\), with \(\Re\Sigma_{\rm ph}^R(\omega)\)
obtained from \cref{eq:ImSigma_subgap_final} via the Kramers--Kronig relation of \cref{eq:ReSigma}.
The figure shows that a finite \(\eta_\Sigma\) rounds off the edge singularity
of \(\Im\Sigma_{\rm ph}^R\) at \(\omega\simeq E_g\) and produces a finite subgap absorption tail.
The corresponding subgap effect on the real part is confined to a narrow frequency window controlled by \(\eta_\Sigma\): moving to lower frequencies below the gap, the finite-broadening curve overlaps again with the zero-damping result.
These subgap effects in the self-energy reflect also at the level of the lower polariton shift.
In \cref{fig:pol_shift_finite_life} we show the lower polariton shift as a
function of \(g/E_g\) for increasing values of \(\eta_\Sigma\). 
At fixed \(\eta_\Sigma>0\) and upon increasing \(g\), the shift is first suppressed but then recovers the  \(\eta_\Sigma=0\) behavior, as the polariton pole is evaluated at frequencies progressively farther from the narrow interval around \(\omega\simeq E_g\) where the finite lifetime most strongly modifies the self-energy. We anticipate that the values of \(\eta_\Sigma\) used in
the figure are intentionally much larger than the cavity-induced spectral broadening estimate derived
below in order to visually magnify the effect.

\subsection{Cavity-induced fermionic spectral broadening from the one-loop Fock self-energy}\label{appsub:fock}

We now estimate the fermionic spectral broadening microscopically. At weak light--matter coupling, the first nonvanishing contribution arises from the one-loop Fock self-energy, which describes a virtual interband scattering process in which a fermion in band \(\lambda\) is temporarily transferred to the opposite band \(-\lambda\) through the interaction with a cavity photon and is then scattered back to its initial state.
We introduce the bare retarded photonic propagator
\begin{align}
D_{\rm ph ,0}^{R}(\omega)
=
\frac{1}{\omega-\omega_0+i\kappa},
\label{eq:Dph_retarded_kappa}
\end{align}
whose linewidth is \(\kappa\).
Since the interaction vertex contains \(\hat{a}+\hat{a}^\dagger\), the object entering the Fock diagram is the propagator of the real quadrature \(\hat{X}=\hat{a}+\hat{a}^\dagger\).
It is convenient to treat it directly through its spectral representation,
\begin{align}
D_X(\omega_\nu)
=
\int_{-\infty}^{\infty}\frac{d\Omega}{2\pi}\,
\frac{B_X(\Omega)}{i\omega_\nu-\Omega},
\label{eq:DX_spectral_appendix}
\end{align}
with bosonic Matsubara frequency \(\omega_\nu\) and spectral density
\begin{align}
B_X(\Omega)
&\equiv
A_{0}(\Omega)-A_{0}(-\Omega),
\label{eq:BX_def_appendix}
\end{align}
where
\begin{align}
A_{0}(\Omega)
&=
-2\,\Im D_{\rm ph ,0}^{R}(\Omega)
=
\frac{2\kappa}{(\Omega-\omega_0)^2+\kappa^2},
\label{eq:Dph_spectral_appendix}
\end{align}
is the bare single-particle spectral function for the complex bosons.
The one-loop Fock self-energy is then
\begin{align}
\Sigma_{\lambda}^{(F)}(\bm{k},\omega_n)
&=
\frac{g(\bm{k})^2}{N}\,
\frac{1}{\beta}\sum_{\omega_\nu}
D_X(\omega_\nu)\,
G_{-\lambda}(\bm{k},\omega_n-\omega_\nu).
\label{eq:Fock_minus_Mats_appendix}
\end{align}
Inserting \cref{eq:DX_spectral_appendix} into \cref{eq:Fock_minus_Mats_appendix} and performing the Matsubara sum using the standard identity
\begin{align}
\frac{1}{\beta}\sum_{\omega_\nu}
\frac{1}{i\omega_\nu-\Omega}\,
\frac{1}{i\omega_n-i\omega_\nu-E}
=
\frac{n_B(\Omega)+n_F(E)}{i\omega_n-E-\Omega},
\end{align}
one obtains
\begin{align}
&\Sigma_{\lambda}^{(F)}(\bm{k},\omega_n)
= \nonumber \\
&\quad\frac{g(\bm{k})^2}{N}
\int_{-\infty}^{\infty}\frac{d\Omega}{2\pi}\,
B_X(\Omega)\,
\frac{n_B(\Omega)+n_F\!\left(E_{-\lambda}(\bm{k})\right)}
{i\omega_n-E_{-\lambda}(\bm{k})-\Omega}.
\end{align}
After analytic continuation \(i\omega_n\to \omega+i0^+\), the exact retarded self-energy reads 
\begin{align}
&\Sigma_{\lambda}^{(F)R}(\bm{k},\omega)
=\nonumber \\
&\quad\frac{g(\bm{k})^2}{N}
\int_{-\infty}^{\infty}\frac{d\Omega}{2\pi}\,
B_X(\Omega)\,
\frac{n_B(\Omega)+n_F\!\left(E_{-\lambda}(\bm{k})\right)}
{\omega-E_{-\lambda}(\bm{k})-\Omega+i0^+}.
\label{eq:Fock_lambda_retarded_appendix}
\end{align}
At zero temperature, \(n_B(\Omega)=-\Theta(-\Omega)\), while
\(n_F(E_-(\bm{k}))=1\) and \(n_F(E_+(\bm{k}))=0\). Therefore, for
\(\lambda=+\),
\begin{align}
n_B(\Omega)+n_F(E_-(\bm{k}))=\Theta(\Omega),
\end{align}
so that the negative-frequency sector of the integral is exactly suppressed.
The retarded self-energy can then be written as
\begin{align}
\Sigma_{+}^{(F)R}(\bm{k},\omega)
=
\frac{g(\bm{k})^2}{N}
\int_{0}^{\infty}\frac{d\Omega}{2\pi}
\frac{B_X(\Omega)}
{z_{+}-\Omega},
\label{eq:Fock_plus_halfaxis_appendix}
\end{align}
with $z_{+}\equiv \omega-E_-(\bm{k})+i0^+ $.
Using \(B_X(\Omega)=A_{0}(\Omega)-A_{0}(-\Omega)\), this becomes
\begin{align}
\Sigma_{+}^{(F)R}(\bm{k},\omega)
&=
\frac{g(\bm{k})^2}{N}
\int_{0}^{\infty}\frac{d\Omega}{2\pi}
\frac{A_{0}(\Omega)}
{z_{+}-\Omega}
\nonumber\\
&\quad
-
\frac{g(\bm{k})^2}{N}
\int_{0}^{\infty}\frac{d\Omega}{2\pi}
\frac{A_{0}(-\Omega)}
{z_{+}-\Omega}.
\label{eq:Fock_plus_split0_appendix}
\end{align}
The first integral can be related to the full spectral representation of the
normal photon propagator. Indeed,
\begin{align}
\int_{0}^{\infty}\frac{d\Omega}{2\pi}
\frac{A_{0}(\Omega)}
{z_{+}-\Omega}
=
D_{\rm ph}^{R}(z_{+})
-
\int_{0}^{\infty}\frac{d\Omega}{2\pi}
\frac{A_{0}(-\Omega)}
{z_{+}+\Omega},
\end{align}
where we reconstructed the full integral over
\(\Omega\in(-\infty,\infty)\) and changed variable
\(\Omega\to-\Omega\) in the negative-frequency sector. Substituting this
identity into \cref{eq:Fock_plus_split0_appendix}, we obtain
\begin{align}
\Sigma_{+}^{(F)R}(\bm{k},\omega)
=
\frac{g(\bm{k})^2}{N}D_{\rm ph}^{R}(z_{+})
+
\delta\Sigma_{+}^{(F)R}(\bm{k},\omega),
\label{eq:Fock_plus_split_appendix}
\end{align}
with
\begin{align}
\delta\Sigma_{+}^{(F)R}(\bm{k},\omega)
&=
-\frac{g(\bm{k})^2}{N}
\int_{0}^{\infty}\frac{d\Omega}{2\pi}
A_{0}(-\Omega)\notag\\
&\times\left[
\frac{1}{z_{+}-\Omega}
+
\frac{1}{z_{+}+\Omega}
\right].
\label{eq:Fock_plus_correction_appendix}
\end{align}
In order to estimate $\eta_{\lambda}$ from \cref{appsub:bubble_finite}, we now
evaluate \cref{eq:Fock_plus_split_appendix} at the HOVHS point \(\bm{k}^*=(\pi/a,0)\), where $E_{\pm}(\bm{k}^*)=\pm {E_g}/{2}$,
and at the resonant frequency \(\omega=\omega_0=E_g\).
For the \(+\) band this gives
\begin{align}
z_{+}^*
\equiv
E_g-E_-(\bm{k}^*)+i0^+
=
\frac{3E_g}{2}+i0^+ .
\end{align}
The first term in \cref{eq:Fock_plus_split_appendix} evaluates to
\begin{align}
\frac{g^2}{N}D_{\rm ph}^R(z_{+}^*)
=
\frac{g^2/N}{E_g/2+i\kappa},
\end{align}
where \(g\equiv g(\bm{k}^*)\). Its magnitude is therefore of order
\begin{align}
\left|\frac{g^2}{N}D_{\rm ph}^R(z_{+}^*)\right|
\propto
\mathcal O\!\left(\frac{g^2}{N E_g}\right),
\qquad
\kappa\ll E_g .
\end{align}
We now show that \(\delta\Sigma_{+}^{(F)R}\) gives only a subleading correction
in the physically relevant narrow-linewidth regime \(\kappa\ll E_g\).
The potentially most delicate part of \(\delta\Sigma_{+}^{(F)R}\) is the neighborhood of
the pole of the first kernel, \(\Omega\simeq z_{+}^*\), because
\((z_{+}^*-\Omega)^{-1}\) is resonant there. We therefore set \(\Omega=z_0+\delta\), with
\(z_0=3E_g/2\) and \(|\delta|\lesssim\kappa\). The new
variable runs over
$\delta\in[-z_0,\infty)$. The point \(\delta=0\) corresponds to the pole of the first kernel,
\(z_+^*-\Omega+i0^+=-\delta+i0^+\). We therefore separate the integral into
a resonant window \(|\delta|<c\kappa\), with \(c\propto\mathcal O(1)\), and the
remaining nonresonant part. The resonant contribution is thus
\begin{align}
I_{\rm res}
=
\int_{-c\kappa}^{c\kappa}
\frac{d\delta}{2\pi}
\frac{A_{0}(-z_0-\delta)}
{-\delta+i0^+}.
\end{align}
 Using
$\frac{1}{-\delta+i0^+}
=
-\mathrm{P.V.}\frac{1}{\delta}
-i\pi\delta(\delta)$,
one finds
\begin{align}
I_{\rm res}
=
-\mathrm{P.V.}\!\int_{-c\kappa}^{c\kappa}
\frac{d\delta}{2\pi}
\frac{A_{0}(-z_0-\delta)}{\delta}
-\frac{i}{2}A_{0}(-z_0).
\end{align}
The imaginary part is therefore controlled by
\begin{align}
A_{0}(-z_0)
=
\frac{2\kappa}{(z_0+\omega_0)^2+\kappa^2}
\propto
\mathcal O\!\left(\frac{\kappa}{E_g^2}\right).
\end{align}
For the principal-value term, since
\(A_{0}(-z_0-\delta)\) varies only on the scale \(E_g\), it can be
expanded as
\begin{align}
A_{0}(-z_0-\delta)
&=
A_{0}(-z_0)
\notag\\&+
\delta\,\partial_\delta A_{0}(-z_0-\delta)\big|_{\delta=0}
+\cdots .
\end{align}
The constant term vanishes in the principal value, while the next term gives
a contribution of order
\begin{align}
\kappa\,\partial_\delta A_{0}(-z_0-\delta)\big|_{\delta=0}
\propto
\mathcal O\!\left(\frac{\kappa^2}{E_g^3}\right).
\end{align}
Thus the resonant part of the correction scales as
\begin{align}
\delta\Sigma_+^{(F)R}(\bm{k}^*,E_g)
\propto
\mathcal O\!\left(
\frac{g^2}{N}
\frac{\kappa}{E_g^2}
\right).
\end{align}
The second denominator in \cref{eq:Fock_plus_correction_appendix} is
nonresonant and gives a contribution of the same or smaller order. By
contrast, the leading term is
\begin{align}
\frac{g^2}{N}D_{\rm ph}^R(z_+^*)
=
\frac{g^2/N}{E_g/2+i\kappa}
\propto
\mathcal O\!\left(\frac{g^2}{N E_g}\right).
\end{align}
Therefore,
\begin{align}
\frac{\delta\Sigma_+^{(F)R}(\bm{k}^*,E_g)}
{\left[g^2/N\right]D_{\rm ph}^R(z_+^*)}
\propto
\mathcal O\!\left(\frac{\kappa}{E_g}\right),
\end{align}
and $\delta\Sigma_+^{(F)R}$ is proved to be subleading in the regime \(\kappa\ll E_g\).
We finally obtain
\begin{align}
\eta_+=-\Im \Sigma_{+}^{(F)R}(\bm{k}^*,E_g)
\approx
\frac{g^2}{N}\,
\frac{\kappa}{(E_g/2)^2+\kappa^2}
\label{eq:ImSigmaPlus_kstar_appendix}
\end{align}

Similarly, for \(\lambda=-\),
\begin{align}
n_B(\Omega)+n_F(E_+(\bm{k}))=-\Theta(-\Omega),
\end{align}
so that only the negative-frequency sector contributes. Changing variable
\(\Omega\to-\Omega\), one obtains
\begin{align}
\Sigma_{-}^{(F)R}(\bm{k},\omega)
=
\frac{g(\bm{k})^2}{N}
\int_{0}^{\infty}\frac{d\Omega}{2\pi}
\frac{B_X(\Omega)}
{z_{-}-\Omega},
\label{eq:Fock_minus_halfaxis_appendix}
\end{align}
with $z_{-}\equiv \omega-E_+(\bm{k})+i0^+ $.
Repeating the same steps gives, to leading order in \(\kappa/E_g\), we obtain
\begin{align}
\eta_- =-\Im \Sigma_{-}^{(F)R}(\bm{k}^*,E_g)
\approx
\frac{g^2}{N}\,
\frac{\kappa}{(3E_g/2)^2+\kappa^2}.
\label{eq:ImSigmaMinus_kstar_appendix}
\end{align}
Using the same realistic setup discussed in \cref{ssec:Feasibility}, we evaluate the Fock self-energy at the coupling \(g\) corresponding to the experimentally achievable pump--cavity strength \(\eta_0\simeq 0.052\,E_{r,c}\) considered there, i.e.~approximately the largest value currently accessible in the cold-atom cavity platforms of interest. We also use the same estimate \(N\simeq 1.1\times 10^{4}\) for the number of unit cells participating in the coupling, obtained in \cref{ssec:Feasibility} from the transverse size of the atomic cloud. With these choices, using the Wannier estimate
$J_x-J_y\simeq0.036\,\eta_0$ for
$\mathcal V_0=5E_{r,L}$ and $\theta\simeq0.424\pi$, the relevant intensive coupling entering
\cref{eq:ImSigmaPlus_kstar_appendix,eq:ImSigmaMinus_kstar_appendix} is
\[
\frac{g}{\sqrt{N}}
=2|J_x-J_y|
\simeq 3.7\times 10^{-3}\,E_{r,c}
\simeq 4.7\times 10^{-3}\,E_g .
\]
Using in addition \(\kappa/h=4.5\,{\rm kHz}\) and $E_g\simeq 3.18 E_{r,L}=h\times 58.7\,{\rm kHz}$, as estimated in \cref{ssec:Feasibility}, we finally obtain
\begin{equation}
\frac{\eta_\Sigma}{E_g}\simeq
\frac{\eta_+}{E_g}
\simeq 6.5\times 10^{-6}.
\label{eqn:final_eta_sigma}
\end{equation}
This estimate shows that the photonic contribution to the fermionic broadening is extremely small with respect to the gap scale \(E_g\) and, based on the results show in \cref{fig:Branch}, essentially negligible at the level of the lower polariton shift.
We note that, in realistic realizations of the model, other experiment-specific decay channels for the atoms (for example, due residual atomic interactions or rare pump-photon absorption processes) could in principle contribute to the fermionic decay and increase $\eta_\Sigma$ with respect to the estimate of \cref{eqn:final_eta_sigma}. In this appendix, we showed however that the \emph{intrinsic} contributions to the spectral broadening in our model are negligible.

\bibliography{bibliography}

\end{document}